\documentclass[journal,transvt]{IEEEtran}
\ifCLASSINFOpdf
\else
\fi
\usepackage{amsmath,amssymb,amsfonts}
\usepackage{subfig}
\usepackage{cite}
\usepackage{algorithm}
\usepackage{graphicx}
\usepackage{enumerate}
\usepackage{bm}
\usepackage{tabularx}
\usepackage{algorithmic}
\usepackage{color}

\newtheorem{theorem}{Theorem}
\newtheorem{definition}{Definition}
\newtheorem{proposition}{Proposition}
\newtheorem{lemma}{Lemma}
\newtheorem{remark}{Remark}
\makeatletter
\newcommand{\multiline}[1]{%
  \begin{tabularx}{\dimexpr\linewidth-\ALG@thistlm}[t]{@{}X@{}}
    #1
  \end{tabularx}
}
\makeatother

\begin{document}
\title{Two-Timescale Resource Allocation for Cooperative D2D Communication: A Matching Game Approach
        \vspace*{-.45cm}  
\thanks{ The material in this paper has been presented in part at 11th International Conference
         		on Wireless Communications and Signal Processing\cite{8927889}.}
 \thanks{Y. Yuan and T. Yang are with the Research Center of Smart Networks and Systems, the Department of Electronic          Engineering, Fudan University, Shanghai, China (e-mail: \{yilingyuan13, taoyang\}@fudan.edu.cn). }
\thanks{Y. Hu is with School of Electronic Information, Wuhan University,  430072 Wuhan, China and ISEK Research Area/Lab, RWTH Aachen University, D-52074 Aachen, Germany (Email: yulin.hu@whu.edu.cn)..}
\thanks{ H. Feng is with the Research Center of Smart Networks and Systems, the Department of Electronic Engineering, Fudan University, Shanghai, China (e-mail: hfeng@fudan.edu.cn). }
\thanks{B. Hu is with the Key Laboratory of EMW Information (MoE), Fudan University, Shanghai, China  (e-mail: bohu@fudan.edu.cn).}
\thanks{This work was supported by the National Key Research and Development Program of China (No. 213), the Shanghai Municipal Natural Science Foundation (No. 19ZR1404700), and China Mobile Chengdu Institute of Research and Development. \emph{Corresponding author: Tao Yang and Bo Hu.}}
}

%\author{\IEEEauthorblockN{Michael Shell}
%\IEEEauthorblockA{School of Electrical and\\Computer Engineering\\
%Georgia Institute of Technology\\
%Atlanta, Georgia 30332--0250\\
%Email: http://www.michaelshell.org/contact.html}
%\and
%\IEEEauthorblockN{Homer Simpson}
%\IEEEauthorblockA{Twentieth Century Fox\\
%\and
%\IEEEauthorblockN{James Kirk\\ and Montgomery Scott}
%\IEEEauthorblockA{Starfleet Academy\\
%San Francisco, California 96678--2391\\


%\author{
%\IEEEauthorblockN{Yiling Yuan\IEEEauthorrefmark{1}, Tao Yang\IEEEauthorrefmark{1},  Yulin Hu\IEEEauthorrefmark{2}, Hui Feng\IEEEauthorrefmark{1} and Bo Hu\IEEEauthorrefmark{1}\IEEEauthorrefmark{3}}
%\IEEEauthorblockA{
%\IEEEauthorrefmark{1} Research Center of Smart Networks and Systems, Fudan University, Shanghai, China\\
%\IEEEauthorrefmark{2} ISEK Research Group, RWTH Aachen University, 52062 Aachen, Germany\\
%\IEEEauthorrefmark{3}Key Laboratory of EMW Information (MoE), Fudan University, Shanghai, China\\
%%Emails: yilingyuan13@fudan.edu.cn, taoyang@fudan.edu.cn, hu@umic.rwth-aachen.de, hfeng@fudan.edu.cn, bohu@fudan.edu.cn
%Emails: \{yilingyuan13, taoyang\}@fudan.edu.cn, hu@umic.rwth-aachen.de, \{hfeng, bohu\}@fudan.edu.cn
%}}

\author{Yiling~Yuan, Tao~Yang,~\IEEEmembership{Member,~IEEE,}
        Yulin~Hu,~\IEEEmembership{Senior Member,~IEEE,}
        Hui~Feng,~\IEEEmembership{Member,~IEEE,}
        and~Bo~Hu,~\IEEEmembership{Member,~IEEE}
        \vspace*{-.35cm}  
         		}% <-this % stops a space

\maketitle

\begin{abstract}
In this paper, we consider a cooperative device-to-device (D2D) communication system, where the D2D transmitters (DTs) act as relays to assist the densified cellular network users (CUs) for transmission quality of service (QoS) improvement. The proposed system achieves a win-win situation, i.e. improving the spectrum efficiency of the CUs that cannot meet their rate requirement while providing spectrum access for D2D pairs. Unlike previous works, to reduce the overhead, we design a novel two-timescale resource allocation scheme, in which the pairing between CUs and D2D pairs is decided at a long timescale and transmission time for CU and D2D pair is determined at a short timescale. Specifically, to characterize the long-term payoff of each potential CU-D2D pair, we investigate the optimal cooperation policy to decide the transmission time  based on the instantaneous channel state \mbox{information (CSI)}. We prove that the optimal policy is a threshold policy which can be achieved via binary search. Since CUs and D2D pairs are self-interested, they are paired only when they agree to cooperate mutually. Therefore, to study the cooperation behaviors of CUs and D2D pairs, we formulate the pairing problem as a matching game, based on the long-term payoff achieved by the optimal cooperation policy of each possible pairing. Furthermore, unlike most previous matching models in D2D networks, we allow transfer between CUs and D2D pairs to improve the performance. To solve the pairing problem, a distributed algorithm is proposed, which converges to an $\epsilon$-stable matching. We show that there is a trade-off between the optimality and the computational complexity of the algorithm. We also analyze the algorithm in terms of the robustness to the unilateral deviation of D2D pairs. Finally, the simulation results verify the efficiency of the proposed matching algorithm.
%We consider a device-to-device (D2D) underlaid cellular network, where each cellular channel can be shared by several D2D pairs and only one channel can be allocated to each D2D pair. We try to maximize the sum rate of D2D pairs while limiting the interference to cellular links. Due to the lack of global information in large scale networks, resource allocation is hard to be implemented in a centralized way. Therefore, we design a novel distributed resource allocation scheme which is based on local information and requires little coordination and communication between D2D pairs. Specifically, we decompose the original problem into two cascaded subproblems, namely channel allocation and power control. The cascaded structure of our scheme enables us to cope with them respectively. Then a two-stage algorithm is proposed. In the first stage, we model the channel allocation problem as a many-to-one matching with externalities and try to find a strongly swap-stable matching. In the second stage, we adopt a pricing mechanism and develop an iterative two-step algorithm to solve the power control problem.
\end{abstract}


% \fi
\IEEEpeerreviewmaketitle

\section{Introduction}
\par Recently,  device-to-device (D2D) communication is recognized as one of the  key technology components in the fifth generation (5G)  and beyond to improve the spectrum efficiency and user experience \cite{Tehrani2014,6970763}. Specifically, proximity D2D users can communicate with each other directly without going through the base station (BS). Taking the advantages of the physical proximity of communicating devices, D2D communication can improve spectrum utilization and energy efficiency, and reduce end-to-end latency. Due to its great potential advantages, the D2D communication has been employed in many applications, such as multicasting \cite{7994919} and relaying \cite{8187632}. There are mainly two ways for D2D pairs to share the cellular spectrum, namely, underlay and overlay D2D \cite{Asadi2014}.  In the underlay D2D communication,  CU and D2D pairs share the same spectrum, which incurs interference to cellular links. Therefore, a set of approaches, such as channel-power allocation \cite{6560489,7308022,7087396} and mode selection \cite{6924793,7174559}, are devoted to alleviating the inter-user interference via applying optimization techniques\footnote{It is {worthy mentioning} that in addition to the optimization techniques, game theoretical models are also adopted to characterize the interaction between BS, UE and D2D users, i.e., referring to \cite{7085929, 6998030, 8078281, 8457531}.}. In contrast, overlay D2D communication allows D2D pairs to occupy dedicated spectrum having been assigned to CUs. Thus, overlay D2D incurs less interference albeit lower spectrum efficiency compared to underlay D2D. Clearly, efficient resource allocation is actually one of the major concerns in the design of such communication systems. For instance, based on stochastic geometry, the authors in \cite{7866013} develop an analytical model for resource allocation in D2D multi-channel overlaying uplink cellular networks, which is helpful to determine the optimal channel allocation to achieve the best network performance. Nevertheless, in both underlay or overlay D2D, the QoS of CUs will be degraded.

\par Meanwhile, the wireless communication network is currently undergoing tremendous changes. Macro-cellular network is evolving to a combination setting, i.e., micro-cell coverage with dense heterogeneous access nodes underlying macro coverage. Thereby the network densification is inevitable, which exacerbates the burden of already strained spectrum resource. In addition, the instant social video can be uploaded or downloaded anytime anywhere, which results in high throughput demand. To ease the shortage of spectrum resource and meet the high data rate demand, emerging ultra-short wavelength communication technologies are developed addressing wide bandwidth, such as millimeter-wave communication. However, these technologies are hindered by high pathloss, shadowing as well as strong interference (especially in the dense access scenario), which limits such transmissions to be only feasible within a short range.

\par Hence, it is challenging to guarantee the QoS requirements of a CU suffering from either a poor channel quality or a strong interference from the neighboring users. 
In this context, cooperative relay is thought as a key technology to tackle this problem, due to its ability in exploiting the spatial diversity gain in the multiuser system \cite{Cao2015MWC}.  Compared to fixed relay stations that incur high expenditure \cite{6775376}, mobile user relaying is an efficient and flexible solution with low cost. Furthermore, it is estimated that the number of mobile devices all over the world, including smartphones and tablets,  will surpass 50 billion with 1 million connections per kilometer square by 2020 \cite{6815890}.
Thus, the massive amount of mobile devices leads to heterogeneous channel conditions, which provides great mobile relaying opportunities for the CUs with poor channels but requiring high rates in the uplink\cite{7248722}. However, mobile terminals serving as relays will consume power. Thus, battery-capacity limited devices are reluctant to provide relay service without any reward. Therefore, it is critical to motivate mobile users to serve as relays \cite{Wu2017TWC}.

\par Combining D2D communication and cooperative relay technology, Wei \emph{et al.} in \cite{Wei2016TVT} propose a D2D-based cooperative network, where mobile devices serve as relays for CUs. However, the work does not consider any  incentive mechanism for mobile devices. Inspired by the idea of spectrum \mbox{leasing \cite{Pantisano2012JSAC}},  a cooperative D2D system is proposed in \cite{Cao2015MWC}, where the DTs act as relays to assist CUs in exchange for the opportunities to use the licensed spectrum.  The rationality behind this system is mainly based on the following two observations. On the one hand, for the uplink transmission, the poor channel condition is the performance limiting  factor for the CUs\cite{Wu2017TWC}. On the other hand, due to the proximity gain of D2D communication \cite{6163598}, D2D devices generally operate at the high signal-to-noise ratio (SNR) regime and thus the bandwidth becomes the performance limiting factor for the D2D pairs \cite{tse2005fundamentals}. Through the cooperation, the QoS of CUs is guaranteed and D2D pairs obtain the transmission opportunities on licensed spectrum. As a result, a win-win situation is achieved, which motivates CUs and D2D pairs to share the spectrum. 

\par Previously, many works \cite{shalmashi2014cooperative,6831680,7544627,8647818,Chen2015WCSP, Yuan2016PIMRC, 7248722, Wu2017TWC} have investigated the cooperative D2D system and have shown the performance gain via the cooperation between D2D pairs and CUs. For example, Shalmashi \emph{et al.} in \cite{shalmashi2014cooperative} aim to minimize the assigned power for cooperation while guaranteeing the performance of CUs. Ma \emph{et al.} propose two superposition coding-based cooperative relaying schemes\cite{6831680}, and further design a contract-based cooperative spectrum sharing mechanism \cite{7544627}. In \cite{8647818}, Gupta \emph{et al.} introduce a best D2D user selection protocol with a novel round robin process to facilitate the fair resource distribution among D2D pairs, and develop a generic mathematical probability model to analyze the performance. However, the results in the above works \cite{shalmashi2014cooperative,6831680,7544627,8647818} are conducted under given user pairs, while  the design of the CU-D2D paring process is ignored. The pairing problem for multiple D2D pairs and CUs is investigated in \cite{Chen2015WCSP, Yuan2016PIMRC, 7248722, Wu2017TWC}. In \cite{Chen2015WCSP}, a coalition formation algorithm is designed to solve the spectrum management problem. In our previous work \cite{Yuan2016PIMRC}, a matching-based resource allocation scheme is proposed to determine the pairing and spectrum allocation. Sun \emph{et al.} in \cite{7248722} propose an energy efficient incentive resource allocation scheme to encourage users to relay data for others with the transmitting resource including time and power as reward. Similarly, the weighted sum energy efficiency of D2D pairs is maximized via a joint D2D relay selection, bandwidth and power allocation in \cite{Wu2017TWC}.

\par However, most works \cite{shalmashi2014cooperative,6831680,7544627,8647818,Chen2015WCSP, Yuan2016PIMRC, 7248722, Wu2017TWC} handle the pairing process between multiple CUs and multiple D2D pairs at a shot timescale (e.g. at LTE scheduling time interval of 1ms), in which {pair switching} may incur heavy signaling overhead, and thus are not practical in large-scale network. With the densification of wireless network and massive connectivity in the future, it is desirable to design a low-overhead resource allocation scheme.

\par To this end, we propose a two-timescale resource allocation scheme in order to reduce the overhead. In particular, the pairing between multiple CUs and multiple D2D pairs is determined at a long timescale. On the other hand, at a short timescale, a cooperation policy allocates the transmission time for D2D link and cellular link based on the instantaneous CSI. Under the proposed scheme, only statistical CSI is required for the pairing problem at the long timescale and the pairing  switches less frequently.  Moreover, at the short timescale, only the instantaneous CSI between every matched CU-D2D pair is acquired by the BS to decide the transmission time. As a result, the signaling overhead is significantly reduced.
%As a result, at short time scale, the BS only needs to collect the instantaneous CSI between every matched CU-D2D pair to decide the transmission time. For the paring problem at long timescale, only statistic CSI is needed. Thus, the signaling overhead can be reduced.

% 待修改
\par Moreover, we develop a matching game based framework to solve the two-timescale resource allocation problem. Specifically, we investigate the optimal cooperation policy for each D2D pair and its potential CU partner to characterize the long-term payoff of this potential pairing. In general,  CUs and D2D pairs may be of self-interest\cite{Song2014}, and thus they can only be paired when they agree to cooperate with each other.  The matching game provides an appropriate framework for such pairing problem \cite{Gu2015MC}. This motivates us to formulate the pairing problem at the long timescale as a one-to-one matching game, which is based on the long-term payoff of each potential CU-D2D pair. %Furthermore, unlike previous one-to-one matching models in D2D networks \cite{Yuan2016PIMRC,Gu2015JSAC,Gu2015MC}, we propose to allow the transfer between CUs and D2D pairs as a performance enhancement.

\par In this paper, we aim to develop a low-overhead resource allocation scheme for cooperative D2D system, where D2D pairs and CUs cooperate mutually via spectrum leasing. In detail, our contributions are summarized as follows.

\begin{itemize}
\item We design a two-timescale resource allocation scheme to reduce overhead. Specifically, the pairing problem is solved at the long timescale. Meanwhile,  the transmission time for CUs and D2D pairs is determined at the short timescale based on the instantaneous CSI between matched users, which is implemented through a cooperation policy.  

\item We investigate the optimal cooperation policy operating at the short timescale to maximize the expected rate of D2D pair while guaranteeing the expected rate requirement of CU. Besides, the optimal policy also plays an important role in bridging the two timescales. In other words, we utilize the performance achieved by the optimal cooperation policy to characterize the long-term payoff of each potential CU-D2D  pair. We prove that the optimal cooperation policy is a threshold policy and can be found via binary search. Moreover, we show that the optimal cooperation policy can also maximize the expected system throughput under some mild conditions.

\item At the long timescale, we formulate the pairing problem between multiple CUs and multiple D2D pairs as a matching game, based on the long-term payoff of each potential matching.  Unlike previous one-to-one matching models in D2D networks \cite{Yuan2016PIMRC,Gu2015JSAC,Gu2015MC,ZhaoICL2017}, we propose to allow the transfer between CUs and D2D pairs as a performance enhancement to solve the inefficiency problem of the stable matching. A distributed matching algorithm is proposed to converge to the $\epsilon$-stable matching. Rigorous analysis shows that there is a trade-off between the optimality and the computational complexity of the algorithm. Another important feature of the algorithm is that it is robust to the unilateral deviation of self-interested D2D pairs. Numerical simulation results verify our argument and  show the efficiency of our matching algorithm.

\end{itemize}

\par The rest of this paper is organized as follows. In Section II, we introduce the system model. We study the optimal cooperation policy in Section III and investigate the pairing problem in Section IV.  Section V gives numerical results. Finally, Section VI concludes this paper.

\par \emph{Notations:} In this paper, $\mathbb{E}\{x\}$ represents the expectation \mbox{of $x$}, and $\mathbf{I}(\cdot)$ denotes the indication function.

\section{System Model}
\par We consider a single cell with a BS denoted by $b$. Because mobile devices are more likely to need help due to their limited  power  budgets, we focus on uplink resource sharing. Assuming there are $M$ CUs suffering from poor channel conditions which could not support their QoS. At the same time, $N$ transmitter-receiver pairs are working in D2D communication mode. There is no dedicated resource allocated for D2D pairs. As a result, D2D pairs serve as relays for CUs in exchange for the access to the cellular channels. In the following, we use $\mathcal{M}=\{1,2,\cdots,M\}$ and $\mathcal{N}=\{1,2,\cdots,N\}$ to denote the sets of CUs and D2D pairs, respectively. For convenience, CU $m$ is denoted by $CU_m$, and D2D pair $n$ is denoted by $DP_n$ and consists of the D2D transmitter $DT_n$ and the D2D receiver $DR_n$. The important symbols of this paper are shown in \mbox{Table. \ref{notations}}.

\begin{table}[!t]
	\footnotesize
	\caption{Summary of important notations}
	\label{notations}
	\centering
	\begin{tabular}{|c|p{6cm}|}
		\hline
		\bf{Symbol} & \bf{Definition}\\
		\hline
		$\mathcal{M}, \mathcal{N}$  & The set of CUs and D2D pairs\\
		\hline
		$h^m_{mb}$  & The channel gain of the cellular link from $CU_m$ to the BS \\
		\hline
		$h^m_{mn}$  & The channel gain of the link from $CU_m$ to $DP_n$ at the channel occupied by $CU_m$\\
		\hline
		$h^m_{nb}$  & The channel gain of the link from $DP_n$ to the BS at the channel occupied by $CU_m$\\
		\hline
		$h^m_{nn}$  & The channel gain of the link from $CU_m$ to the BS at the channel occupied by $CU_m$\\
		\hline
		$\alpha_{mn}$  & The time allocation factor for $CU_m$ and  $DP_n$\\
		\hline
		$\pi$  & The cooperation policy\\
		\hline
		$\bf{X}$ & Assignment matrix for D2D pairs and CUs \\
		\hline
		$\mu$ & One-to-one mapping which matches the D2D pair with one CU \\
		\hline
		$\bf{p}$ & The price vector of CUs \\
		\hline
		$\Phi$  & The matching between the CUs and D2D pairs\\
		\hline
		$\theta_m(\Phi)$  & The utility of $CU_m$ under the matching $\Phi$\\
		\hline
	    $\delta_n(\Phi)$  & The utility of $DP_n$ under the matching $\Phi$\\
		\hline
		
	\end{tabular}
\end{table}

\par The time domain is divided into frames of fixed length. We assume that the pairs of D2D users remain the same during the entire frame\footnote{The assumption is reasonable, which are thought as the representative use-cases of D2D communications in cellular networks \cite{Asadi2014}, especially for high data rate services (e.g. video sharing, gaming). On the other hand, we want to point out that it is also interesting to consider the scheme design without this assumption, i.e. for low data rate services, and this  will be considered in our future work.}. Each frame consists of $T_s$ subframe. Assuming the channel gain remains constant in each subframe and changes over different subframes. Besides, we assume that the channel gains across different subframes of the same frame are i.i.d. and follow a known distribution. At the channel occupied by $CU_m$, the instantaneous channel gains of the cellular link from $CU_m$ to BS, the link from $CU_m$ to $DT_n$, the link from $DT_n$ to BS and the D2D link from $DT_n$ to $DR_n$ are represented as $h^m_{mb}, h^m_{mn}, h^m_{nb}, h^m_{nn}$, respectively.

\begin{figure}[!t]
\centering
\includegraphics[width=3in]{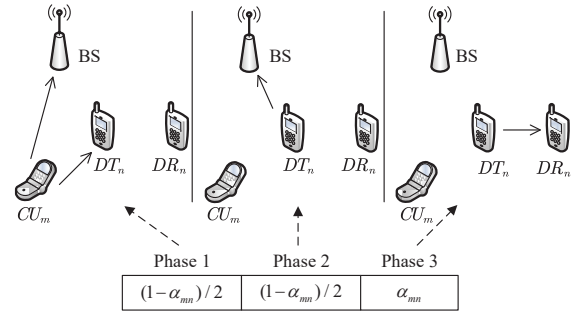}
\captionsetup{font={small}}
\caption{Subframe structure for cooperation.}
\label{systemModel}
\end{figure}

\par We assume that each CU is assisted by at most one D2D pair since it has been shown that a single relay can achieve the full diversity gain \cite{Kadloor2010TWC}. As depicted in Fig. \ref{systemModel}, the normalized subframe is divided into three phases when $DP_n$ cooperates with  $CU_m$\footnote{For simplicity, we assume the transmission direction of  D2D pair is fixed during the entire frame and D2D transmitter acts as a relay for CUs. In fact, our proposed scheme can be applied to a more general scenario, where  the transmission direction may change and both D2D transmitter and D2D receiver can be selected as a relay.}. The first two phases both last $\frac{1-\alpha_{mn}}{2}$ and are used for the relay transmission of  $CU_m$, where  $\alpha_{mn}\in\mathcal{A}\triangleq[0,1]$ is referred as the \emph{time allocation factor}. Specifically, $CU_m$ broadcasts its data with power $P_c$ to the BS and $DT_n$ at first. Then, $DT_n$ forwards the received data to the BS with power $P_d$. Besides, the first two phases can also be used for the cellular link of  $CU_m$ when the cellular link has better performance. The last phase lasts $\alpha_{mn}$ and is used for D2D link, where $DT_n$ communicates with $DR_n$ with power $P_d$. 

\par The rate of  $CU_m$ in the cellular link is 
\begin{equation}
r^C_{m}=\ln\left(1+\frac{P_ch^m_{mb}}{N_0}\right),
\end{equation}
where $N_0$ denotes the noise power. In this paper, we take the decode-and-forward relay scheme as an example. Thus, when an entire subframe is used for a relay transmission of $CU_m$, this relay transmission rate is given by
\begin{equation}
r^R_{mn} = \frac{1}{2}\min\left\{\ln\left(\!1\!+\!\frac{P_ch^m_{mn}}{N_0}\!\right),\ln\left(\!1\!+\!\frac{P_ch^m_{mb}}{N_0}\!+\!\frac{P_dh^m_{nb}}{N_0}\!\right)\right\}.
\end{equation}
Note that if the quality of the cellular link is better, the first two phases can also be used for the cellular link of  $CU_m$. As a result, the achieved rate of  $CU_m$ during the entire subframe is represented as
\begin{equation}
R^C_{mn}(\alpha_{mn})=(1-\alpha_{mn})\max\left\{r^C_m,r^R_{mn}\right\}.
\end{equation}
For convenience, we define $r^C_{mn}\triangleq\max\left\{r^C_m,r^R_{mn}\right\}$, which is the maximum rate of $CU_m$ during the subframe, and can be achieved when the entire subframe is used for the cellular transmission.

\par At the same time, the rate of $DP_n$ during the entire subframe is given as
\begin{equation}
R^D_{mn}(\alpha_{mn})=\alpha_{mn}\ln\left(1+\frac{P_dh^m_{nn}}{N_0}\right)\triangleq\alpha_{mn}r^D_{mn},
\end{equation}
where $r^D_{mn}$ is the maximum rate of $DP_n$ during the subframe and can be achieved when the entire subframe is spent for D2D transmission.

\par  Thus, we have two variables to determine: pairing between multiple CUs and multiple D2D pairs, and time allocation factor for each CU-D2D pair. To this end, we propose a matching game based framework to determine these two variables at two different timescales. Specifically, based on the instantaneous CSI, a cooperation policy decides the time allocation factor for each CU-D2D pair at each subframe (i.e. at the short timescale). We aim at characterizing  the long-term payoff of each potential CU-D2D pair via the optimal cooperation policy. Then, following the characterizations of this long-term payoff, we use the matching game with transfer to decide the pairing for each frame (i.e. at the long timescale).  In the following two sections, we will study these two subproblems, respectively.

\begin{remark} 
	To gather the instantaneous global CSI, certain overhead of the proposed two-timescale scheme  should be spent. In particular, the overhead of the proposed two-timescale scheme is $\mathcal{O}(\min\{M,N\})$ while the overhead of a one-timescale scheme is $\mathcal{O}(MN)$. Therefore, in comparison to the one-timescale scheme, the overhead for CSI collection is reduced via the proposed two-timescale design, especially for large-scale networks (with large values of $M$ and $N$). Moreover, the two-timescale scheme has less frequent pairing switch. This advantage further alleviates the signaling overhead.
\end{remark}

\section{Optimal Cooperation Policy}
\par In this section, we investigate the optimal cooperative policy for each CU-D2D pair. Without loss of generality,   we  study a pair of  $CU_m$ and $DP_n$. Define the state $\mathbf{r}_{mn}\triangleq(r^C_{mn},r^D_{mn})$ representing the instantaneous CSI. The set of all the possible states is denoted by $\mathcal{R}_{mn}$. The cooperation policy determines the time allocation factor $\alpha_{mn}$ according to the current state $\mathbf{r}_{mn}$. Mathematically, the cooperation policy is a function $\pi:\mathcal{R}_{mn}\rightarrow\mathcal{A}$. Thus, given the state $\mathbf{r}_{mn}$, the rate of $DP_n$ and   $CU_m$ can be represented as $\pi(\mathbf{r}_{mn})r^D_{mn}$ and $\left(1-\pi(\mathbf{r}_{mn})\right)r^C_{mn}$, respectively. 

\par Our objective is to find the optimal policy maximizing the expected rate of the D2D pair while  guaranteeing the QoS of the CU. Therefore, the optimization problem is formulated as  
\begin{subequations}
\label{equ5}
\begin{alignat}{2}
  \max_{\pi}\quad &\mathbb{E}_{\mathbf{r}_{mn}}\left\{\pi(\mathbf{r}_{mn})r^D_{mn}\right\} \label{equ5:a}\\
  \text{s.t.}\quad & \mathbb{E}_{\mathbf{r}_{mn}}\left\{\left(1-\pi(\mathbf{r}_{mn})\right)r^C_{mn}\right\}\geq r_{th},\label{equ5:b}
\end{alignat}
\end{subequations}
where $r_{th}$ is the minimum rate requirement for the CU and constraint (\ref{equ5:b}) is used to guarantee the QoS of  the CU. In fact, if $T_s\gg 1$, the objective function (\ref{equ5:a}) and the left-hand side of constraint (\ref{equ5:b}) are good approximations of the average rates of $DP_n$ and $CU_m$ over $T_s$ subframes, respectively. In this section, all the expectations are taken over the random \mbox{variable $\mathbf{r}_{mn}$}. For brevity, we omit the subscript $\mathbf{r}_{mn}$ in the following discussion. 

\par Next, we investigate the structure of the optimal cooperation policy in the following theorem. 

\begin{theorem}[Structure of Optimal Policy]
\label{thm1}
If problem (\ref{equ5}) is feasible, the optimal policy $\pi^*$ is given by
\begin{equation}
\label{equ6}
\pi^*(\mathbf{r}_{mn})=\begin{cases}
0,&\quad\lambda^*r^C_{mn}>r^D_{mn},\\
\alpha^*,&\quad\lambda^*r^C_{mn}=r^D_{mn},\\
1,&\quad\lambda^*r^C_{mn}<r^D_{mn},
\end{cases}
\end{equation}
where
\begin{align}
\lambda^*&=\min\left\{\lambda|\mathbb{E}\{r^C_{mn}\mathbf{I}(\lambda r^C_{mn}\geq r^D_{mn})\}\geq r_{th}\right\},\\
\alpha^*   &=\frac{r_{th}-\mathbb{E}\{r^C_{mn}\mathbf{I}(\lambda^* r^C_{mn}> r^D_{mn})\}}{\mathbb{E}\{r^C_{mn}\mathbf{I}(\lambda^* r^C_{mn}= r^D_{mn})\}} \label{equ2-8}.
\end{align}
\end{theorem}

\begin{remark}Theorem \ref{thm1} indicates that the optimal policy can be a threshold policy, which makes decisions based on the ratio of $r^D_{mn}$ to $r^C_{mn}$. Equality (7) implies that $\lambda^*$ is the minimum threshold which can satisfy constraint (5b). In other words, for any threshold policy having the similar form as shown in (\ref{equ6}) and satisfying constraint (\ref{equ5:b}), the corresponding threshold of the policy is definitely no smaller than $\lambda^*$. According to the policy shown in (\ref{equ6}), the expected rate of $CU_m$ is $\mathbb{E}\{r^C_{mn}\mathbf{I}(\lambda^* r^C_{mn}> r^D_{mn})\} + \alpha^*\mathbb{E}\{r^C_{mn}\mathbf{I}(\lambda^* r^C_{mn}= r^D_{mn})\}$. Note that problem (\ref{equ5}) suggests that as long as constraint (\ref{equ5:b}) is guaranteed, $\alpha^*$ should be as small as possible. As a result, $\alpha^*$ is obtained as shown in (\ref{equ2-8}).  
\end{remark}

\begin{IEEEproof}
We construct the Lagrangian for problem (\ref{equ5}) as follows.
\begin{equation}
\label{equ2-9}
\begin{split}
\mathcal{L}(\pi,\lambda)&=\mathbb{E}\left\{\pi(\mathbf{r}_{mn})r^D_{mn}\right\}+\lambda\left(\mathbb{E}\{(1-\pi(\mathbf{r}_{mn}))r^C_{mn}\}\!-\!r_{th}\right)\\
&=\mathbb{E}\left\{\pi(\mathbf{r}_{mn})(r^D_{mn}-\lambda r^C_{mn})\right\}+\lambda\mathbb{E}\{r^C_{mn}\}-\lambda r_{th}, 
\end{split}
\end{equation}
where $\lambda$ is the Lagrange multiplier associated with constraint (\ref{equ5:b}). For a fixed $\lambda$, it is easy to find out that the following \mbox{policy $\hat{\pi}_{\lambda}$}, which is given in (\ref{solutionOfLagarange}), can maximize the Lagrangian $\mathcal{L}(\pi,\lambda)$.
\begin{equation}
\label{solutionOfLagarange}
\hat{\pi}_{\lambda}(\mathbf{r}_{mn})=\begin{cases}
0,&\quad\lambda r^C_{mn}>r^D_{mn},\\
\alpha^*,&\quad\lambda r^C_{mn}=r^D_{mn},\\
1,&\quad\lambda r^C_{mn}<r^D_{mn}.
\end{cases}
\end{equation}

\par The Lagrange dual function is given by $g(\lambda) = \max_{\pi}\mathcal{L}(\pi,\lambda)$. Thus, substituting (\ref{solutionOfLagarange}) to $g(\lambda)$, we have
\begin{equation}
\label{dual}
\begin{split}
g(\lambda) 
=& \mathcal{L}(\hat{\pi}_{\lambda},\lambda)\\
=& \mathbb{E}\left\{\mathbf{I}(\lambda r^C_{mn}\!<\!r^D_{mn})(r^D_{mn}\!-\!\lambda r^C_{mn})\right\}+\lambda\mathbb{E}\{r^C_{mn}\}-\lambda r_{th}.
\end{split}
\end{equation}

\par In the following, we show that $\lambda^*$ minimizes the Lagrange dual function.

\par Assuming $\Delta\lambda>0$. Then, using (\ref{dual}), we have
\begin{align*}
&g(\lambda^*+\Delta\lambda)-g(\lambda^*)\\
&\qquad=\Delta\lambda\mathbb{E}\{r^C_{mn}\}-\Delta\lambda r_{th}-\Delta\lambda\mathbb{E}\left\{r^C_{mn}\mathbf{I}(\lambda^*r^C_{mn}\!<\! r^D_{mn}) \right\}\\
&\qquad\quad-\mathbb{E}\Big\{(\lambda^*\!+\!\Delta\lambda)r^C_{mn}\mathbb{I}(\lambda^*r^C_{mn}\!\leq\! r^D_{mn\!}<\!(\Delta\lambda\!+\!\lambda^*)r^C_{mn}\Big\}\\
&\qquad\quad+\mathbb{E}\Big\{r^D_{mn}\mathbf{I}(\lambda^*r^C_{mn}\!\leq\! r^D_{mn\!}<\!(\Delta\lambda\!+\!\lambda^*)r^C_{mn}\Big\}\\
&\qquad\geq\Delta\lambda\mathbb{E}\{r^C_{mn}\}-\Delta\lambda\mathbb{E}\left\{r^C_{mn}\mathbf{I}(\lambda^*r^C_{mn}\!<\! r^D_{mn}) \right\}-\Delta\lambda r_{th}\\
&\qquad=\Delta\lambda\mathbb{E}\left\{r^C_{mn}\mathbf{I}(\lambda^*r^C_{mn}\!\geq\! r^D_{mn}) \right\}-\Delta\lambda r_{th}\\
&\qquad\geq 0,
\end{align*}
where the last inequality is based on the definition of $\lambda^*$.

\par On the other hand, it is easy to verify that $\hat{\pi}_{\lambda^*}(\mathbf{r}_{mn})\leq\hat{\pi}_{\lambda^*-\Delta\lambda}(\mathbf{r}_{mn})$. Consequently, using the definition of $\lambda^*$, we have the following inequalities
\begin{align*}
&g(\lambda^*)-g(\lambda^*-\Delta\lambda)\\
&\qquad\leq -\Delta\lambda\mathbb{E}\left\{\hat{\pi}_{\lambda^*}(\mathbf{r}_{mn})r^C_{mn}\right\}+\Delta\lambda\mathbb{E}\{r^C_{mn}\}-\Delta\lambda r_{th}\\
&\qquad=  -\Delta\lambda\mathbb{E}\{r^C_{mn}\mathbf{I}(\lambda^*r^C_{mn}\!<\!r^D_{mn})\}\!+\!\Delta\lambda\mathbb{E}\{r^C_{mn}\}\!-\!\Delta\lambda r_{th}\\
&\qquad=  \Delta\lambda\mathbb{E}\{r^C_{mn}\mathbf{I}(\lambda^*r^C_{mn}\geq r^D_{mn})\}-\Delta\lambda r_{th}\\
&\qquad\leq  0.
\end{align*}

\par Thus, we conclude that $\lambda^*$ is a solution to the dual problem $\min_{\lambda\geq 0}g(\lambda)$. Therefore, we have
\begin{equation*}
P^*\overset{(a)}\leq g(\lambda^*)=\mathcal{L}(\pi^*,\lambda^*)\overset{(b)}=\mathbb{E}\left\{\pi^*(\mathbf{r}_{mn})r^D_{mn}\right\}\overset{(c)}\leq P^*,
\end{equation*}
where $P^*$ is the optimal value of problem (\ref{equ5}). Inequality (a) is due to the duality gap.  \mbox{Equality (b)} is based on the fact that the policy $\pi^*$ makes \mbox{constraint  (\ref{equ5:b})} hold with equality. Since $\pi^*$ is a feasible solution to problem (\ref{equ5}), we obtain the inequality (c). 

\par Therefore, we conclude that $\pi^*$ is an optimal cooperation policy.
\end{IEEEproof}

\par Recall that Theorem \ref{thm1} implies that the optimal policy can be a threshold policy. Besides, this theorem also shows that this optimal policy will allocate the entire subframe for D2D transmission (i.e. $\alpha_{mn}=1$) or for cellular transmission (i.e. $\alpha_{mn}=0$) in most cases. As a result, such optimal policy enables efficient implementation in practice. 

\par Furthermore, the term $\mathbb{E}\left\{r^C_{mn}\mathbf{I}(\lambda r^C_{mn}\geq r^D_{mn})\right\}$ increases with increasing $\lambda$. Therefore,  the binary search could be adopted to find the threshold $\lambda^*$ with low complexity, which is depicted in Algorithm 1.

\begin{algorithm}[!t]
\caption{Binary search to find the optimal cooperation policy}
 {%\fontsize{8pt}{0.85\baselineskip}\selectfont
 \begin{algorithmic}[1]
 \STATE  Set $\lambda_u=\lambda_{max}$ and $\lambda_l=0$;
 \WHILE{$|\lambda_u-\lambda_l|>\epsilon_0$}
 \STATE  $\lambda_{mid}=(\lambda_u+\lambda_l)/2$;
 \IF{$\mathbb{E}\{r^C_{mn}\mathbf{I}(\lambda_{mid}r^C_{mn}\geq r^D_{mn})\}\leq r_{th}$}
 \STATE Set $\lambda_l=\lambda_{mid}$;
 \ELSE
 \STATE Set $\lambda_u=\lambda_{mid}$;
 \ENDIF
 \ENDWHILE
 \STATE Set $\lambda^*=\lambda_{mid}$;
 \STATE Calculate $\alpha^*$ according to (\ref{equ2-8}).
 \end{algorithmic}}
 \end{algorithm}

\par Another interesting fact about the optimal policy $\pi^*$ is that $\pi^*$ can also maximize the expected weighted sum rate of CUs and D2D pairs under some mild conditions. The intuition behind this fact is that the channel quality of the D2D link is usually better than that of cellular link. The following proposition states this fact formally.

\begin{proposition}
\label{prop1}
Consider the following optimization problem:
\begin{subequations}
\label{equ2-15}
\begin{alignat}{3}
  \max_{\pi}\quad &\mathbb{E}\left\{\pi(\mathbf{r}_{mn})r^D_{mn}+\eta \left(1-\pi(\mathbf{r}_{mn})\right)r^C_{mn}\right\}\label{equ2-15:a}\\
  \text{s.t.}\quad & \mathbb{E}\left\{\left(1-\pi(\mathbf{r}_{mn})\right)r^C_{mn}\right\}\geq r_{th},\label{equ2-15:b}\\
  \quad & \mathbb{E}\left\{\pi(\mathbf{r}_{mn})r^D_{mn}\right\}\geq r_{min,D}.\label{equ2-15:c}
\end{alignat}
\end{subequations}
If the above problem is feasible and satisfies the following condition
\begin{equation}
\label{equ2-16}
\text{Pr}\left\{r^D_{mn}>\eta r^C_{mn}\right\}>1-\frac{r^2_{th}}{\mathbb{E}\left\{(r^C_{mn})^2\right\}},
\end{equation}
then the policy $\pi^*$ shown in (\ref{equ6}) is also the optimal solution to problem (\ref{equ2-15}).
\end{proposition}
\begin{IEEEproof}
See Appendix A.
\end{IEEEproof}

\par Note that problem (\ref{equ2-15}) is a general form of problem (\ref{equ5}). Hence, Proposition \ref{prop1} claims that $\pi^*$ is also the optimal solution to this more general problem when (\ref{equ2-16}) is satisfied. The condition means that the channel quality of the D2D link should be sufficiently better than that of cellular link. Due to the short distances of D2D links, if the weight $\eta$ is selected appropriately (such as $\eta=1$), the condition often holds in practice.

\section{Matching Game for Pairing Problem}
\par In this section, we study the pairing problem. The assignment is represented as a binary matrix $\mathbf{X}_{M\times N}=[x_{mn}]$, where $x_{mn} = 1$ implies that $CU_m$ and $DP_n$ are matched. We define $u_{mn}\triangleq\mathbb{E}\left\{\pi^*(\mathbf{r}_{mn})r^D_{mn}\right\}$ if \mbox{problem (\ref{equ5})} is feasible. In the case of infeasibility, we set $u_{mn} = -1$. Thus, $ u_{mn}$ characterizes the long-term payoff of $DP_n$ when it cooperates with $CU_m$. Besides, when $u_{mn}\geq 0$, we call $DP_n$ being \emph{acceptable} to $CU_m$. On the contrary, we call $DP_n$ being \emph{unacceptable} to $CU_m$ when $u_{mn}< 0$. The objective is to maximize the long-term average weighted sum rate of D2D pairs, which is formulated as the following problem.
\begin{subequations}
\label{equ11}
\begin{alignat}{3}
  \max_{\mathbf{X}}\quad & V(\mathbf{X})=\sum_{n\in\mathcal{N}}w_n\sum_{m\in\mathcal{M}}x_{mn}u_{mn}&& \label{equ11:a}\\
  \text{s.t.}\quad & \sum_{n\in\mathcal{N}}x_{mn}\leq 1, \quad\forall m\in\mathcal{M},&&\label{equ11:b}\\
  					& \sum_{m\in\mathcal{M}}x_{mn}\leq 1,\quad\forall n\in\mathcal{N},&&\label{equ11:c}\\
  					& x_{mn}\in\{0,1\},\quad\forall m\in\mathcal{M},\forall n\in\mathcal{N},&&\label{equ11:d}
\end{alignat}
\end{subequations}
where $w_n$ is the weight associated with $DP_n$ and determined by the QoS requirement of $DP_n$. Constraint (\ref{equ11:b}) makes sure that each CU is relayed by at most one DT. Define $v_{mn}\triangleq w_nu_{mn}$. Thus, $v_{mn}$ is the value of the coalitions $\{CU_m,DP_n\}$. The vector $\mathbf{v}_n=(v_{1n},\cdots,v_{Mn})$ is referred to as the value vector of $DP_n$. Due to the limited battery capacity, each D2D pair relays at most one CU \cite{Wu2017TWC}, which is represented in constraint (\ref{equ11:c}). Note that $v_{mn}<0$ when $DP_n$ is unacceptable to $CU_m$. Therefore, the CUs will be only matched with acceptable D2D pairs. 
\subsection{Matching Game with Transfer}
\par Originally stemmed from economics \cite{Roth1990Two}, the matching theory provides a framework to tackle the  problem of pairing players in two distinct sets, based on each player's individual preference. Since the CUs and D2D pairs are usually self-interested, we use the matching game to characterize the cooperation between CUs and D2D pairs in the pairing problem. Moreover, Theorem 1 implies that CU is indifferent over the acceptable D2D pairs while D2D pairs may have strict preferences over CUs. Therefore, if the achieved long-term rate is used to determine the preferences of users, an inefficient stable matching will be obtained finally. To this end, we allow transfer between D2D pairs and CUs to improve the performance. Such model is called matching game with transfer\cite{Bayat2016MSP} and also referred as to assignment game\cite{Roth1990Two}. Specifically, each CU has a price charged to its matched partner. In fact, the price of a CU indicates the willingness of D2D pairs to cooperate with it. 

\begin{definition}
A \emph{one-to-one mapping} $\mu$ is a function from $\mathcal{M}\cup\mathcal{N}$ to $\mathcal{M}\cup\mathcal{N}\cup\{0\}$ such that $\mu(m)=n$ if and only if $\mu(n)=m$, and $\mu(m)\in\mathcal{M}\cup\{0\}$, $\mu(n)\in\mathcal{N}\cup\{0\}$ for $\forall m\in\mathcal{M},\forall n\in\mathcal{N}$.
\end{definition}

\par Note that $\mu(x)=0$ means that the user $x$ is unmatched in $\mu$. The above definition implies that a one-to-one mapping matches a user on one side to the one on the other side unless the user is unmatched. In other words, a mapping $\mu$ defines a feasible solution to problem (\ref{equ11}). In the following, we will introduce the price into the matching model.
\begin{definition}
A \emph{matching} is defined as $\Phi=(\mu,\mathbf{p})$, where $\mu$ is a one-to-one mapping, $\mathbf{p}=(p_1,p_2,\cdots,p_M)$ is the price vector of CUs and $p_m\geq 0,\forall m \in \mathcal{M}$. Moreover, if $\mu(m)=0$, then $p_m=0$.
\end{definition}

\par We denote the utilities of $CU_m$ and $DP_n$ as $\theta_m$ and $\delta_n$, respectively. Thus, given a matching $\Phi=(\mu,\mathbf{p})$, $\theta_m$ and $\delta_n$ is represented as
\begin{align}
\theta_m(\Phi)&=p_m,\\
\delta_n(\Phi) &= v_{\mu(n)n}-p_{\mu(n)}.
\end{align}
Here, we let $p_0=0$ and $v_{0n}=0,\forall n\in\mathcal{N}$ for convenience.

\par In matching theory, the concept of \emph{stability} is important. Since the price is usually discrete for exchange among users in practice, we introduce the $\epsilon$-stable matching\footnote{A more general definition of $\epsilon$-stable matching can be found in \cite{Hamza2017JSAC}.} as follows. 
\begin{definition}
Given $\epsilon\geq0$, a matching $\Phi$ is \emph{$\epsilon$-stable}, if and only if the following two conditions are satisfied:
\begin{enumerate}[(1)]
\item $\theta_m(\Phi)\geq 0,\delta_n(\Phi)\geq 0$, for $\forall m\in\mathcal{M},\forall n\in\mathcal{N}$;
\item $\theta_m(\Phi)+\delta_n(\Phi)\geq v_{mn}-\epsilon$, for $\forall m\in\mathcal{M},\forall n\in\mathcal{N}$.
\end{enumerate}
\end{definition}
Condition (1) is called \emph{individual rationality condition} and reflects that a user may remain unmatched if the cooperation is not beneficial. Condition (2) implies that there is no CU-D2D pair $(m,n)$ such that they can form a new matching, where both of them  have higher utilities while one of them improves its utility by at least $\epsilon$.

\par The following theorem presents the lower bound of the achievable D2D pairs' sum rate in the $\epsilon$-stable matching.    
\begin{theorem}
\label{the2-2}
Given an $\epsilon$-stable matching  $\Phi$, its associated assignment $\mathbf{X}_{\Phi}$ satisfies the following inequality
\begin{equation}\label{the2}
V(\mathbf{X}_{\Phi})\geq V(\mathbf{X}^*)-\epsilon\min\{M,N\}, 
\end{equation}
where $\mathbf{X}^*$ is the optimal solution to problem  (\ref{equ11}).
\end{theorem}
\begin{IEEEproof}
Consider the following problem
\begin{subequations}
\label{equ15}
\begin{alignat}{3}
  \min_{\theta,\delta}\quad & U(\theta,\delta)=\sum_{m\in\mathcal{M}}\theta_m+\sum_{n\in\mathcal{N}}\delta_n&& \label{equ15:a}\\
  \text{s.t.}\quad & \theta_m\geq0,\delta_n\geq0, \quad\forall m\in\mathcal{M},\forall n\in\mathcal{N},&&\label{equ15:b}\\
  					&\theta_m+\delta_n\geq v_{mn}-\epsilon, \quad\forall m\in\mathcal{M},\forall n\in\mathcal{N},&&\label{equ15:c}
\end{alignat}
\end{subequations}
where $\theta=(\theta_1,\cdots,\theta_M)$ and $\delta=(\delta_1,\cdots,\delta_N)$. Denote  problem (\ref{equ15}) as problem $P(\epsilon)$. It can be shown that the dual problem of problem $P(0)$ is problem (\ref{equ11}), where $x_{mn}$ is the Lagrange multiplier associated with constraint (\ref{equ15:c})\cite{Roth1990Two}.

\par Define $\theta(\Phi)=(\theta_1(\Phi),\cdots,\theta_M(\Phi))$ and $\delta(\Phi)=(\delta_1(\phi),\cdots,\delta_N(\Phi))$. Thus, $(\theta(\Phi),\delta(\Phi))$ is a feasible solution to problem $P(\epsilon)$ and problem $P(0)$. Note that $V(\mathbf{X})$ is the Lagrange dual function of problem $P(0)$. Therefore, for the optimal assignment $\mathbf{X}^* = [x^*_{mn}]$, we have 
\begin{align*}
V(\mathbf{X}^*)\leq &U(\theta (\Phi ),\delta (\Phi ))+\!\!\!\!\!\! \!\sum_{m \in {\cal M},n \in {\cal N}}\!\!\!\!\!\!\! {x_{mn}^*\left( {{v_{mn}} - {\theta _m}(\Phi ) - {\delta _n}(\Phi )} \right)}\\
\leq &U(\theta (\Phi ),\delta (\Phi ))+\!\!\!\!\!\! \!\sum_{m \in {\cal M},n \in {\cal N}} \!\!\!\!\!\! \!\epsilon x_{mn}^*\\
\leq &U(\theta (\Phi ),\delta (\Phi ))+\epsilon\min\{M,N\}\\
= &V(\mathbf{X}_{\Phi})+\epsilon\min\{M,N\}.
\end{align*}
According to the definition of Lagrange dual function, the first equality can be obtained. Because $(\theta(\Phi),\delta(\Phi))$ satisfies constraint (\ref{equ15:c}), the second equality holds. The third equality is due to the fact that $\sum_{m \in {\cal M},n \in {\cal N}}x_{mn}^*\leq\min\{M,N\}$. Therefore, (\ref{the2}) holds.
\end{IEEEproof}

\begin{algorithm}[!t]
\caption{Distributed Matching Algorithm (DMA)}
 {%\fontsize{8pt}{0.85\baselineskip}\selectfont
 \begin{algorithmic}[1]
 \renewcommand{\algorithmicrequire}{\textbf{Initialization:}}
 %\Require
 \STATE Set $t=1, p_m=\beta_m^t=0,\mu^0(m)=0,\forall m\in\mathcal{M}$;
  \renewcommand{\algorithmicrequire}{\textbf{D2D Pairs' Proposals:}}
 \REQUIRE
 \STATE  Broadcast the price requirement vector $\boldsymbol{\beta}^t=({\beta}^t_1,\beta^t_2,\cdots,\beta^t_M)$;
 \FOR{each unmatched D2D pair $n\in\mathcal{N}$}
 \STATE  Determine its demand $m=D_n(\bm{\beta}^t)$;
  \STATE  { If $m\neq 0$, $DP_n$ proposes to $CU_m$ \mbox{($g^t_{mn} = 1$)}.  Otherwise, $DP_n$ does not proposes ($g^t_{mn} = 0,\forall m\in\mathcal{M}$) and $\mu^t(n)=0$;}
\ENDFOR
 \renewcommand{\algorithmicrequire}{\textbf{CUs' Decision-Making:}}
\REQUIRE
 \FOR{Each CU $m\in\mathcal{M}$}
 \IF{$\sum_{n\in\mathcal{N}}g^t_{mn}\!=\!0$, $\sum_{n\in\mathcal{N}}g^{t-1}_{mn}>0$ and $\mu(m)=0$}
  \STATE  Set $\mu^t(m)=n^*$, where $n^*=random(\{n|g^{t-1}_{mn}\!=\!1\})$;
 \STATE Set $p_m=\beta^{t-1}_m$ and $\beta^{t+1}_m = \beta^{t}_m$;
  \STATE  Set $g^t_{m^*n^*}=0$, where $m^* = D_{n^*}(\bm{\beta}^t)$;
 \ENDIF
 \ENDFOR
 \FOR {Each CU $m\in\mathcal{M}$}
 \IF {$\sum
 \limits_{n\in\mathcal{N}}g^t_{mn}\!=\!1$, and either $\mu^{t-1}_m\!=\!0$ or $p_m\!<\!\beta^t_m$ are satisfied }
  \STATE Set $\mu^{t}(m)=n^*$, where $g^t_{mn^*}=1$;
  \STATE Set $p_m=\beta^t_m$ and $\beta^{t+1}_m=\beta^t_m$;
% \Else
% \State Set $\mu^t(m)=0$;
% \State Set $\beta^{t+1}_m=\beta^t_m+\epsilon$;
\ELSIF{$\sum_{n\in\mathcal{N}}g^t_{mn}\geq1$}
   \STATE  Set $\mu^t(m)=0$;
   \STATE  If $n\!\neq\!0$ and $p_m\!=\!\beta^t_m$ where $n\!=\!\mu^{t-1}(m)$, set $g^t_{mn}\!=\!1$;
  \STATE  Set $\beta^{t+1}_m=\beta^t_m+\epsilon$;
 \ELSE
  \STATE Set $\beta^{t+1}_m=\beta^t_m$;
 \ENDIF
 \ENDFOR
  \STATE  $t\leftarrow t+1$;
 \STATE  Go to step 3 until there is no proposal in the current loop.
 \end{algorithmic}}
 \end{algorithm}

\subsection{Distributed Matching Algorithm}
 \par  The algorithm to find an $\epsilon$-stable matching is depicted in Algorithm 2. In the following, we give a brief description of the algorithm during $t$-th iteration.
 \par At first, the price requirement vector $\bm{\beta}^t$ will be broadcast, and $\beta^t_m$ represents the minimum price has to pay if D2D pair wants to propose to $CU_m$ at the current iteration. Then, each unmatched $DP_n$ selects its favorite CU according to $D_n(\bm{\beta}^t)$, where the demand function is represented as {\footnote{We assume there is no tie in (\ref{demandfunction}), which can be realized by adding some sufficiently small positive number to each value $v_{mn}$.}}
 \begin{equation}
 \label{demandfunction}
 D_n(\bm{\beta}^t)=\begin{cases}
 arg\max\limits_ {m\in\mathcal{M}}(v_{mn}-\beta_m^t),&\max\limits _{m\in\mathcal{M}}(v_{mn}-\beta_m^t)\geq0 ,\\
 0,&\text{otherwise.}
 \end{cases}
 \end{equation}
 
 \par In the CUs' decision-making stage, the CUs decide if they want to match with the D2D pairs. There are four cases for each CU. The first case (step 8-12) is that $CU_m$ is unmatched and receives no proposals after increasing its price requirement, but has received proposals from multiple D2D pairs in the previous iteration. Then, $CU_m$ will select randomly one of those D2D pairs to be matched with and set the price as $p_m=\beta^{t-1}_m$. The second case (step 15-17) is that $CU_m$ receives one proposal, and meanwhile, it is unmatched or matched with price $p_m<\beta^t_m$. In other words, only one D2D pair wants to be matched with $CU_m$ with price $\beta^t_m$.  As a result, $CU_m$ will be matched with that D2D pair and set the price as $p_m=\beta^t_m$. The third  case (step 18-21) is that there are multiple D2D pairs (including the current partner of $CU_m$) wanting to be matched with $CU_m$ with the price $\beta^t_m$. Then,  $CU_m$ will increase its price requirement by $\epsilon$ and become unmatched, where $\epsilon> 0$ is the price-step.  In the other situations \mbox{(step 23)}, $CU_m$ will remain the price requirement and do nothing else. The convergence of the DMA is given in the following theorem.
\begin{theorem}
The DMA converges to an $\epsilon$-stable matching.
\end{theorem}

\begin{IEEEproof}
At first, we show that the algorithm converges to a matching. Note that $\beta^t_m$ is non-decreasing. Moreover, it can be found that $\beta^t_m\leq\epsilon+\max_{n\in\mathcal{N}}v_{mn}$. Therefore, the algorithm will converge in finite steps. Use $\Phi=(\mu,\mathbf{p})$ to denote the final result. It is easy to verify that once a CU has received a proposal,  the CU will have a partner in $\mu$. Thus, the prices of the CUs unmatched in $\mu$ are zero. Therefore, $(\mu,\mathbf{p})$ is a matching.

\par In the following, we prove that $(\mu,\mathbf{p})$ is $\epsilon$-stable by contradiction. 

\par Suppose there exists a CU-D2D pair $(m,n)$ such that $\theta_m+\delta_n< v_{mn} -\epsilon$. Assume $\mu(m)=n'$ and $\mu(n)=m'$. Thus, we can find a price $p'$ such that $p'\geq p_m+\epsilon$ and $v_{mn}-p'>v_{m'n}-p_{m'}$. So, we have
\begin{equation}
\label{equ14}
v_{mn}-p_m>v_{mn}-p'>v_{m'n}-p_{m'}.
\end{equation}
According to the algorithm, (\ref{equ14}) implies that $DP_n$ must have proposed to $CU_m$. Therefore, there must exist an iteration, denoted by $\tau$-th iteration, where the first case happens for $CU_m$. Specifically, $CU_m$ receives multiple proposals with $\beta^{\tau-1}_m=p_m$ at $(\tau-1)$-th iteration, and receives no proposal with $\beta^{\tau}_m=p_m+\epsilon$  at  $\tau$-th iteration. Furthermore, no D2D pairs propose to $CU_m$ afterward. As a result, we have the following inequalities
\begin{equation}
v_{mn}-p'\leq v_{mn}-\beta^\tau_m<v_{m'n}-p_{m'},
\end{equation}
which is inconsistent with (\ref{equ14}). 
\par Besides, it is easy to verify $(\mu,\mathbf{p})$ satisfies the individual rationality condition. Therefore, we conclude that $(\mu,\mathbf{p})$ is $\epsilon$-stable.
\end{IEEEproof}

\begin{proposition}
The computational complexity of the DMA is $\mathcal{O}(\frac{MNV_{max}}{\epsilon})$, where $V_{max}=\max_{m,n} v_{mn}$.
\end{proposition}
\begin{IEEEproof}
The price requirement vector $\bm{\beta}$ is non-decreasing in the algorithm. There are at most $\mathcal{O}(\frac{MV_{max}}{\epsilon})$ distinct price requirement vectors. Note that at least one D2D pair's proposal is rejected until convergence. Furthermore, once one D2D pair is rejected by $CU_m$ at price requirement $\beta_m$, then it will not propose to $CU_m$ with $\beta_m$ afterward. Therefore, the computational complexity of the algorithm is $\mathcal{O}(\frac{MNV_{max}}{\epsilon})$ in the worst case.
\end{IEEEproof}

\subsection{Truthfulness of the DMA}
In the DMA, the weight $w_n$ is the private information of $DP_n$. Without knowing $w_n$ globally, the DMA can still get a suboptimal solution to problem (\ref{equ11}). However, after receiving the price requirements of CUs, each D2D pair may be a decision-maker and choose any other CUs except the CU determined by the demand function. Thus, a question will arise: is such deviation behavior beneficial to the D2D pair? If the answer is NO, we call that the D2D pairs are truthful and the DMA satisfies truthfulness property. In fact, the truthfullness is an important property in mechanism design \cite{shoham2008multiagent}, which has been studied widely in wireless communication. For instance, a truthful reverse-auction is designed for computation offloading in \cite{8489932} and a truthful two-tier architecture is developed for spectrum auction in \cite{8769864}. In the following, we will show that the DMA is robust to such deviation.

\par Note that above deviation behaviors can be restated as an equivalent form: the D2D pair deviates from its true value vector and chooses the CU by calculating the demand function but based on the deviated value vector. We will focus on this equivalent form in the following analysis. Besides, the $\epsilon$-stable matching obtained by the DMA is denoted by $\Phi$ for convenience.  

\begin{lemma}
\label{lemma2-2}
{In the matching $\Phi$ obtained by DMA under the announced value vector $\{\mathbf{v}_n\}_{n\in\mathcal{N}}$,} the utility of any D2D pair $n\in\mathcal{N}$ satisfies the following inequalities
\begin{align}
\delta_n(\Phi)&\!\leq \!V(\mathcal{M},\mathcal{N})\!-\!V(\mathcal{M},\mathcal{N\!}\setminus\!\{n\})+4C_1\epsilon, \label{equ2-33}\\
\delta_n(\Phi)&\!\geq \!V(\mathcal{M},\mathcal{N})\!-\!V(\mathcal{M},\mathcal{N}\!\setminus\!\{n\})-(C_1\!+\!C_2\!+\!1)\epsilon,\label{equ2-32}
\end{align}
where $C_1=\min\{M,N-1\}$ and $C_2=\min\{M,N\}$. Besides, $V(\mathcal{M}_1,\mathcal{N}_1)$ is the value of the optimal assignment  {with respect to the value vectors $\{\mathbf{v}_n\}_{n\in\mathcal{N}}$} when the sets of CUs and D2D pairs is $\mathcal{M}_1$ and $\mathcal{N}_1$ respectively.
\end{lemma}
\begin{IEEEproof}
See Appendix B.
\end{IEEEproof}

\par 
{If  $DP_n$ is matched with $CU_m$ in $\Phi$ obtained by DMA,} it is easy to verify the following inequalities:
\begin{align}
V(\mathcal{M},\mathcal{N})&\geq  v_{mn}+V(\mathcal{M}\setminus\{m\},\mathcal{N}\setminus\{n\}),\label{equ2-41-1}\\
V(\mathcal{M},\mathcal{N})&\leq  v_{mn}+V(\mathcal{M}\setminus\{m\},\mathcal{N}\setminus\{n\})+C_2\epsilon.\label{equ2-41-2}
\end{align}
Hence, combining (\ref{equ2-41-1}), (\ref{equ2-41-2}) and Lemma \ref{lemma2-2}, we further obtain the lower bound and the upper bound of $\delta_n(\Phi)$, which are given in (\ref{equ2-42-1}) and (\ref{equ2-42-2}).
\begin{equation}
\begin{split}
\delta_n(\Phi)&\geq V(\mathcal{M}\setminus\{m\},\mathcal{N}\setminus\{n\})-V(\mathcal{M},\mathcal{N}\setminus\{n\})  \\
&\qquad+v_{mn}- (C_1+C_2+1)\epsilon,\label{equ2-42-1}
\end{split}
\end{equation}
\begin{equation}
\begin{split}
\delta_n(\Phi) &\leq V(\mathcal{M}\setminus\{m\},\mathcal{N}\setminus\{n\})-V(\mathcal{M},\mathcal{N}\setminus\{n\})  \\
&\qquad+v_{mn} + (4C_1+C_2)\epsilon,\label{equ2-42-2}
\end{split}
\end{equation}

\par {Furthermore, using $p_m = v_{mn}-\delta_{n}(\Phi)$, the lower bound and the upper bound of $p_m$ can be given as follows
\begin{equation}
\label{equ2-43-1}
\begin{split}
p_m & \geq V(\mathcal{M},\mathcal{N}\setminus\{n\}) - V(\mathcal{M}\setminus\{m\},\mathcal{N}\setminus\{n\})\\ 
&\qquad-(4C_1+C_2)\epsilon,\\
\end{split}
\end{equation}
\begin{equation}
\label{equ2-43-2}
\begin{split}
p_m &  \leq  V(\mathcal{M},\mathcal{N}\setminus\{n\}) - V(\mathcal{M}\setminus\{m\},\mathcal{N}\setminus\{n\})\\ 
&\qquad+(C_1+C_2+1)\epsilon.
\end{split}
\end{equation}
}

\par The critical observation from (\ref{equ2-43-1}) and (\ref{equ2-43-2})   is that the upper bound and the lower bound of the price do not depend on any values $v_{mn}$ of $DP_n$  {and  only depends on the final matching and the value vectors of other D2D pairs.} This fact implies the robustness of the DMA, which is stated in Theorem \ref{thm2-4}.

\begin{theorem}
\label{thm2-4}
In the DMA, each D2D pair can improve its utility by at most $(8\min\{M,N\}+1)\epsilon$ via unilaterally deviating from its true value vector.
\end{theorem}
\begin{IEEEproof}
We prove the theorem holds for a random D2D pair $n\in\mathcal{N}$.  {The true value vector of D2D pairs is denoted by $\{\mathbf{v}_n\}_{n\in\mathcal{N}}$. Define $V(\mathcal{M}_1,\mathcal{N}_1)$ is the value of the optimal assignment with respect to the values $\{v_{mn}\}_{m\in\mathcal{M}_1,n\in\mathcal{N}_1}$.} 
\par Assume  $DP_n$ obtains the payoffs $\delta_n$ and $\delta'_n$ under the true value vector and under the deviated value vector, respectively. There are only two cases.

\par In the first case, $DP_n$ is matched with $CU_m$ under the true value vector. Assume $DP_n$ is matched with $CU_k$ and get the utility $\delta'_n = v_{kn} - p_k$ under the deviated true value. Thus,, we have
\begin{align*}
\delta'_n 
&\leq v_{kn}+V(\mathcal{M}\setminus\{k\},\mathcal{N}\setminus\{n\})-V(\mathcal{M},\mathcal{N}\setminus\{n\})\\
&\qquad+(4C_1+C_2)\epsilon \\
&\leq V(\mathcal{M},\mathcal{N})-V(\mathcal{M},\mathcal{N}\setminus\{n\})+(4C_1+C_2)\epsilon\\
&\leq v_{mn}+V(\mathcal{M}\setminus\{m\},\mathcal{N}\setminus\{n\})-V(\mathcal{M},\mathcal{N}\setminus\{n\})\\
&\qquad+(4C_1+2C_2)\epsilon\\
& \leq \delta_n +  (5C_1+3C_2+1)\epsilon\\
& \leq \delta_n + (8\min\{M,N\}+1)\epsilon.
\end{align*}
{The first inequality follows the lower bound given in (\ref{equ2-43-1}). The second inequality is based on the optimality of $V(\mathcal{M}, \mathcal{N})$. Using the sub-optimality of the $\epsilon$-stable matching given in Theorem \ref{the2-2}, the third inequality can be derived. The fourth inequality is based on (\ref{equ2-42-1}).} Besides, it is easy to verify the theorem holds when $DP_n$ is unmatched after it deviates from its true value vector.

\par In the second case,  $DP_n$ is unmatched under the true value vector. Then, the inequality $V(\mathcal{M},\mathcal{N})\leq V(\mathcal{M},\mathcal{N}\setminus\{n\})+C_2\epsilon$ holds. Thus, we have
\begin{align*}
\delta'_n &\leq V(\mathcal{M},\mathcal{N})-V(\mathcal{M},\mathcal{N}\setminus\{n\})+(4C_1+C_2)\epsilon\\
& \leq C_2\epsilon + (4C_1+C_2)\epsilon\\
& \leq \delta_n +(8\min\{M,N\}+1)\epsilon,
\end{align*}
where the first inequality can be obtained similarly as the first case.

\par Therefore, the theorem is verified.
\end{IEEEproof}

\begin{remark} 
	In practice, decision-making ability may incur certain computational cost. From this point of view, Theorem \ref{thm2-4} actually indicates an interesting consequence that if $\epsilon$ is sufficiently small, each D2D pair will not deviate from its true value since the utility improvement is trivial in this situation. Thus, the DMA can achieve suboptimal matching even if the value vector is unknown globally and the D2D pairs may be intelligent to deviate from its true value vector.
\end{remark}

\section{Simulation Results}
In this section, the performance of the proposed algorithm is investigated through simulations. The instantaneous channel gain used in the simulation is $h=\xi L^{-\gamma}$, where $\xi$ is the fast fading gain with exponential distribution, $\gamma=4$ is the pathloss exponent and $L$ is the distance between transmitter and receiver. In our simulations, we consider the scenario where the BS is deployed in the cell center while the radius of the cell is set to 500 m. The CUs are distributed uniformly at the cell edge. Meanwhile, the D2D pairs are uniformly distributed in the area with a distance of 200 m to 400 m from the BS. Without loss of generality, all the weights $w_n (n=1,\cdots,N)$ are set to 1. Other configuration parameters are given in \mbox{Table. \ref{parametersTable}}. Without a specific illustration, all the results are averaged over 1000 simulations.

\begin{table}[!t]
\footnotesize
\caption{Configuration Parameters}
\renewcommand{\arraystretch}{1.1}
\label{parametersTable}
\centering
\begin{tabular}{|c|c|}
\hline
\bf{Parameters} & \bf{Value}\\
\hline
Power noise ($N_0$) & -100 dBm\\
\hline
Transmit power of CU ($P_c$) & 20 mW\\
\hline
Transmit power of DT ($P_d$) & 20 mW\\
\hline
Distance of D2D link & Uniformly distributed in $[10, 30]$ m\\
\hline
Minimum rate requirement ($r_{th}$) & 1.8 bps/Hz\\
\hline
Number of CUs ($M$) & $15\sim30$\\
\hline
Number of D2D pairs ($N$) & $5\sim40$\\
\hline
Price-step ($\epsilon$) & $1/8\sim 8$\\
\hline
\end{tabular}
\end{table}

\subsection{The Property of the DMA}

\begin{figure}[!t]
\centering
\captionsetup{font={small}}
\includegraphics[width=3.5in]{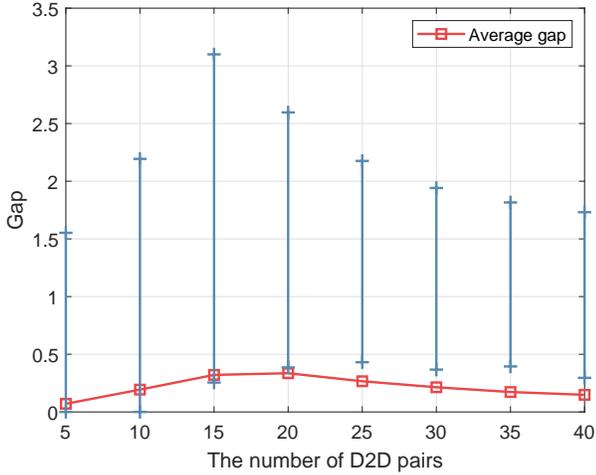}
\caption{The gap $|V(\mathcal{M},\mathcal{N})-V(\mathcal{M},\mathcal{N}\setminus\{n\})-\delta_n|$ with different number of D2D pairs, where $M=15,\epsilon=1$.}
\label{gap}
\end{figure}

\par To verify the upper bound and the lower bound given in Lemma \ref{lemma2-2}, Fig. \ref{gap} investigates the gap $|V(\mathcal{M},\mathcal{N})-V(\mathcal{M},\mathcal{N}\setminus\{n\})-\delta_n|$ with different number of D2D pairs. The vertical blue lines present the range of the maximum gap for each experiment with the given number of D2D pairs. The figure shows that the maximum gap is less than $3.5\epsilon$ and the average gap is even less than $0.5\epsilon$. Therefore, the simulation results imply that the bound given in  Lemma \ref{lemma2-2} is loose in practice. Furthermore, these results also indicate that in practice, the DMA has stronger robustness to the unilateral deviation of D2D pairs, in comparison to our analysis results given in Theorem \ref{thm2-4}.

\begin{figure}[!t]
\centering
\captionsetup{font={small}}
\includegraphics[width=3.5in]{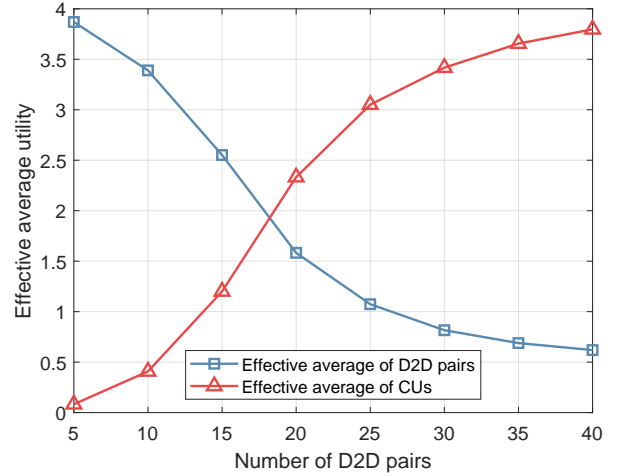}
\caption{The effective average utilities of D2D pairs and CUs versus the number of D2D pairs, where $M=15$.}
\label{AvgUtility}
\end{figure}

\par Fig. \ref{AvgUtility} presents the effective average  utilities (EAU) of CUs and D2D pairs versus the number of D2D pairs. The EAU of CUs is defined as $\text{EAU} = \frac{\text{Sum of CUs' utilities}}{\text{Number of matched CUs}}$. The EAU of D2D pairs can be defined in a similar way. It can be observed that with the increasing number of D2D pairs, the EAU of CUs increases while the EAU of D2D pairs decreases. The rationality behind this is that when $N$ is small, there is a strong competition among CUs on acquiring the relay services from D2D pairs. Therefore, the prices of CUs are low and each matched D2D pair has high utility. On the other hand, when there exist a large number of D2D pairs, the available CUs become the scarce resource. As a result, each D2D pair has to pay a higher price for the transmission opportunities on the cellular channels.

\subsection{Comparison with Other Matching Algorithms}
\par
\begin{figure}[!t]
\centering
\captionsetup{font={small}}
\includegraphics[width=3.5in]{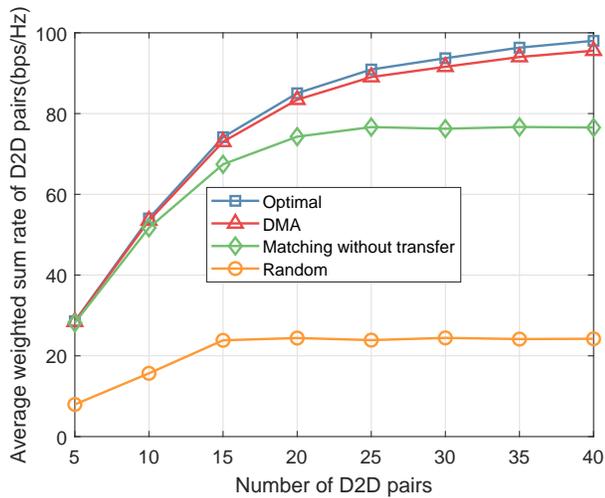}
\caption{The average weighted sum rate of D2D pairs with different matching algorithms versus the number of D2D pairs, where $M=15$ and $\epsilon=1$.}
\label{SumRateD2D}
\end{figure}

\begin{figure}[!t]
\centering
\captionsetup{font={small}}
\includegraphics[width=3.5in]{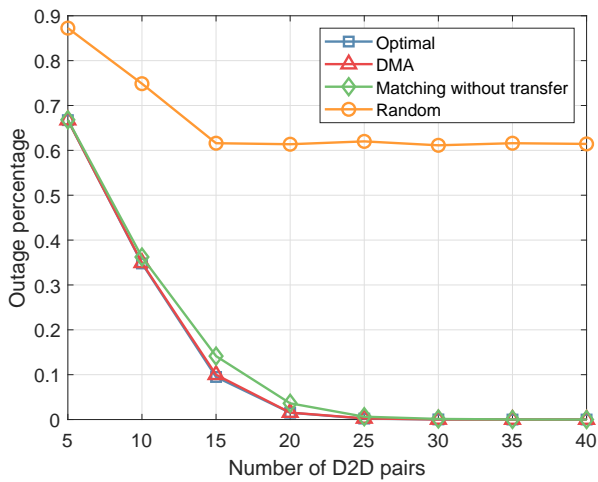}
\caption{The outage percentage of CUs with different matching algorithms versus the number of D2D pairs, where $M=15$ and $\epsilon=1$.}
\label{Outage}
\end{figure}

 In this part, to evaluate the performance of the DMA, we compare it with the following algorithms: i) the \emph{optimal} matching is the optimal solution to problem  (\ref{equ11}) and can be found by Hungarian algorithm  {but global information is needed;} ii) the  \emph{matching without transfer} that adopts matching game for pairing problem (\ref{equ11}), but there is no transfer between CUs and D2D pairs, i.e. the prices of CUs are zero; iii) a \emph{random} algorithm matching the D2D pairs and CUs randomly. The comparison results are provided in Fig. \ref{SumRateD2D} and Fig. \ref{Outage}.
% In the following, we will compare the proposed scheme with other schemes. The \emph{optimal} scheme adopts the optimal solution to problem (\ref{equ11}). The scheme, referred as to \emph{matching without transfer}, adopts matching game for pairing problem (\ref{equ11}) but does not allow transfer between CUs and D2D pairs, i.e. the prices of CUs are zero. The \emph{random} scheme matches the D2D pairs and CUs randomly.

\par In Fig. \ref{SumRateD2D}, we compare the performance of different algorithms in terms of the average weighted sum rate of D2D pairs. This figure shows that the DMA achieves near-optimal performance. Besides, owning to allowing transfer between CUs and D2D pairs, the DMA outperforms the matching without transfer, especially in the large $N$ regime. It also can be observed that the gain is small in the small $N$ regime. This is due to the fact the prices of CUs are close to zero when $N$ is small (which is consistent with the results in Fig. \ref{AvgUtility}). Thus, these two algorithms are almost the same in this situation. Moreover, the matching without transfer only matches the CUs with their acceptable partners. Therefore, this scheme obtains better performance than the random scheme.

\par Fig. \ref{Outage} presents the outage percentage of CUs under different algorithms. The outage refers to the case where the rate requirement of a CU is not satisfied. Compared with the random algorithm, the rest three algorithms achieve significantly better performance. In particular, these three algorithms have the similar outage percentage. The explanation is as follows. The outage never happens if each CU is matched with an acceptable D2D pair. Since all the three algorithms only match the CUs with their acceptable D2D pairs, the price has little impact on the outage percentage. Furthermore, when $N\geq20$, the outage percentage of the DMA is close to zero. It implies that the DMA improves the performance of CUs greatly. On the contrary, the outage percentage of the random algorithm is larger than 60\%, which indicates that it is essential to have an efficient pairing between CUs and D2D pairs. Finally, it is worth noting that all the CUs cannot satisfy their rate requirements without cooperating with D2D pairs in our simulation settings. Therefore, it can be found that the cooperation scheme improves the performance of CUs significantly even the random matching is applied.

\subsection{Comparison with One-timescale Scheme}
\begin{figure}[!t]
	\centering
	\subfloat[The average weighted sum rate of D2D pairs versus the number of D2D pairs.]{\includegraphics[width=3.5in]{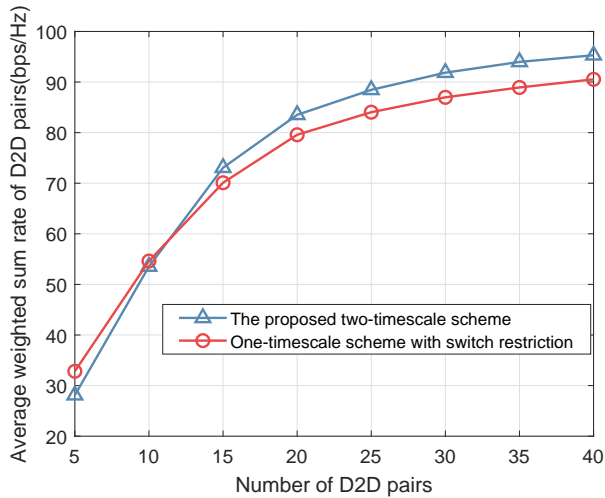}}
	\hfill
	\subfloat[The outage percentage of CUs versus the number of D2D pairs.]{\includegraphics[width=3.5in]{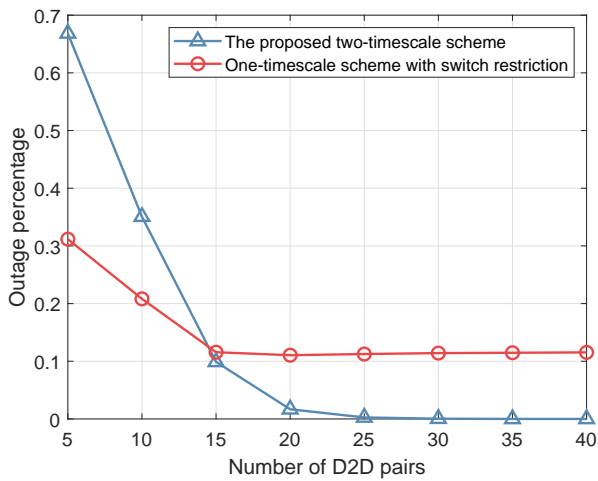}}
	\captionsetup{font={small}}
	\caption{Comparison with one-timescale scheme, where $M=15$ and $\epsilon=1$.}
	\label{oneTimeScale}
\end{figure}

In the following, we compare the proposed two-timescale scheme with a one-timescale scheme. Generally, one-timescale resource allocation schemes, which determine the pairing and the time allocation factor {at a short time-scale}, can achieve a better performance but often need much more CU-D2D matching switches than the two-timescale schemes. In other words, the two-timescale schemes reduce the switching overhead greatly at the cost of certain performance degradation. To investigate the trade-off between performance and switching overhead under these two kinds of resource allocation schemes, we consider one-timescale scheme with switch restriction as a benchmark. In this one-timescale scheme, optimal time allocation factor is determined at each subframe under the constraint that at most two CU-D2D matching  can be changed between adjacent subframes\footnote{On the one hand, allowing more than two changes (e.g. three or more) would increase the complexity dramatically. On the other hand, no matching change is needed within the entire frame in the two-timescale scheme. Therefore, we only consider the restriction where at most two matching changes are allowed.}. The results are provided in Fig. \ref{oneTimeScale}, which are averaged over 1000 frames with different topologies while there are 1000 subframes in each frame.

\par
It can be observed that the one-timescale scheme with switch restriction performs better when the number of D2D pairs is relatively small. The rationale is given as follows. With less D2D pairs, only a few times of switching is needed to achieve the optimal performance at each subframe, which means that the switch restriction has little influence on the performance of one-timescale scheme. In such case, the one-timescale scheme with switch restriction outperforms the two-timescale one. On the other hand, when there are a large number of D2D pairs, the number of allowed switches becomes insufficient, which is actually the bottleneck of the one-timescale scheme. Besides, the one-timescale scheme only focuses on the current subframe while the two-timescale scheme takes the entire frame into account. Consequently, as observed in Fig. \ref{oneTimeScale}, the proposed two-timescale scheme is more preferred when the number of D2D pairs is relatively large, which is usually the case in practice. Hence, the results in \mbox{Fig. \ref{oneTimeScale}} confirms that the proposed design achieves better trade-off between the performance and the switching overhead in practical cellular networks supporting a relatively larger number of D2D pairs.

\subsection{Impact of Price-step $\epsilon$}
\begin{figure}[!t]
\centering
\captionsetup{font={small}}
\includegraphics[width=3.5in]{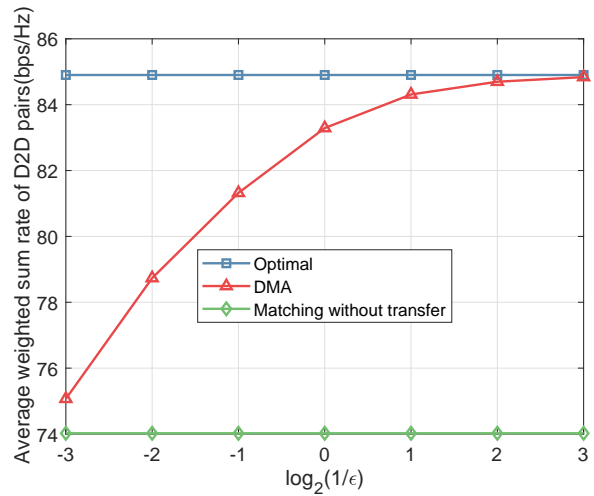}
\caption{The average weighted sum rate of D2D pairs with different $\epsilon$, where $M=N=15$.}
\label{epsilonOpt}
\end{figure}

We investigate the impact of price-step $\epsilon$ in this part. Fig. \ref{epsilonOpt} presents the average weighted sum rate of D2D pairs versus price-step $\epsilon$. On the one hand, with the decreasing $\epsilon$, the performance of the DMA increases and converges to the optimal performance gradually, which complies with Theorem \ref{the2-2}. On the other hand, if $\epsilon$ is so large that it is close to or even larger than the maximum of $v_{mn}$, then there are a few of D2D pairs willing to be matched with the CU with price $\epsilon$. As a result, in the final matching obtained by the DMA, most of the CUs have price 0. Thus,  as shown in Fig. \ref{epsilonOpt}, the DMA has little performance gain over the matching without transfer in large $\epsilon$ regime.

\par
\begin{figure}[!t]
\centering
\captionsetup{font={small}}
\includegraphics[width=3.5in]{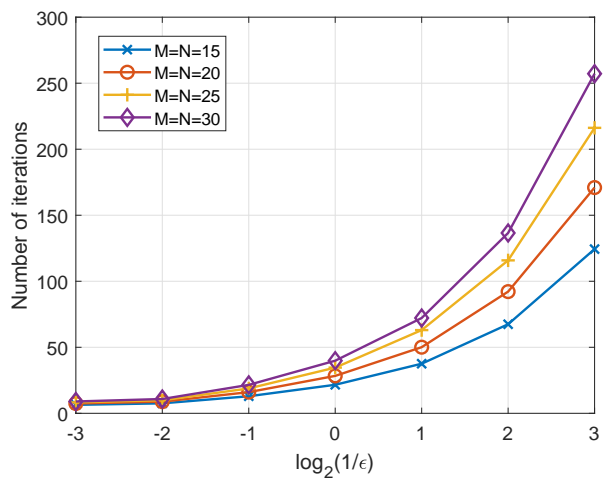}
\caption{The number of iterations with different $\epsilon$.}
\label{epsilonComplex}
\end{figure}

 Fig. \ref{epsilonComplex} investigates the impact of $\epsilon$ on the complexity of the DMA. It can be observed that the complexity of the DMA increases with the decreasing $\epsilon$, which matches with Proposition 2. Besides, the complexity of the DMA does not increase dramatically for larger networks. Combining the results shown in Fig. \ref{epsilonOpt} and  Fig. \ref{epsilonComplex}, it can be found that increasing $\epsilon$ could reduce the complexity of the DMA at the expense of larger performance loss. Therefore, an appropriate price step should be selected to balance the trade-off between the optimality and complexity. For example, in our settings, when $M=N=15$, setting $\epsilon=1$ can achieve near-optimal performance with low complexity.

\subsection{Impact of User Mobility}
\begin{figure}[!t]
	\centering
	\subfloat[The average weighted sum rate of D2D pairs over time with different speed.]{\includegraphics[width=3.5in]{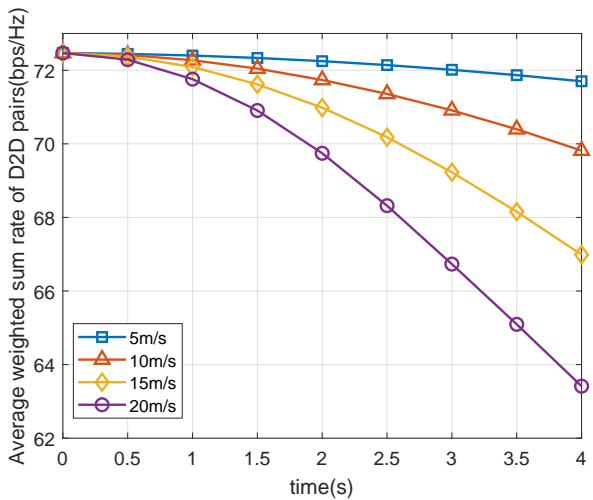} \label{sumRateD2DMobility}}
	\hfill
	\subfloat[The outage percentage of CUs over time with different speed.]{\includegraphics[width=3.5in]{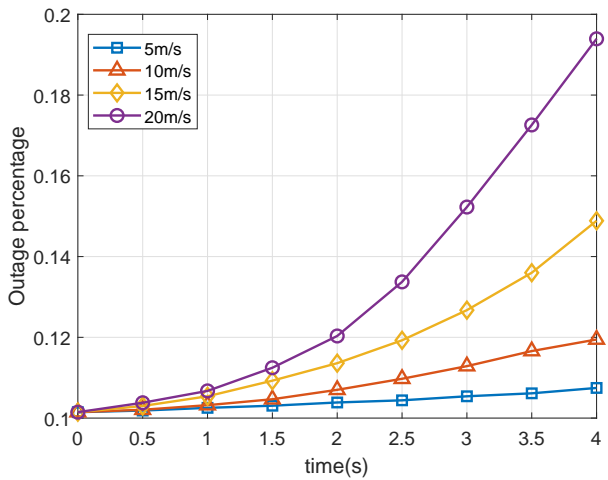}\label{outageCUMobility}} 
	\captionsetup{font={small}}
	\caption{The impact of user mobility, where $N=M=15$ and $\epsilon=1$.}
	\label{Mobility}
\end{figure}
At last, the impact of user mobility on the performance of our scheme is investigated. We consider a random mobility model where CUs and D2D pairs move at random directions with fixed speed. In particular, the matching is determined by DMA at the beginning and remains unchanged afterwards. The simulation results are shown in Fig. \ref{Mobility}. The performance of D2D pairs is presented in Fig. \ref{sumRateD2DMobility}. It can be found that the D2D pairs have the almost same performance during long period (e.g. 4s) in low mobility case. For the high mobility case, e.g. 20m/s, only about $5\%$ performance loss is observed for D2D pairs if the same matching is adopted in the period lasting 2s. Fig. \ref{outageCUMobility} evaluates the influence of the mobility in terms of the performance of CUs. The similar observations can be obtained: low mobility has little impact on the performance of CUs while the performance loss is small for high mobility case even if the matching changes every two seconds. Hence, the results in Fig. \ref{Mobility} confirm that the proposed scheme is insensitive to the users' mobility.

\section{Conclusion}
In this paper, we have investigated a cooperative D2D communication system, where D2D pairs and CUs cooperate with each other via spectrum leasing. We have provided a low-overhead design for the system by proposing a two-timescale resource allocation scheme, in which the pairing between CUs and D2D pairs is decided at the long timescale while time allocation factor is determined at the short timescale. Specifically, to characterize the long-term payoff of each potential CU-D2D pair, we investigated the optimal cooperation policy to decide the time allocation factor. We analyzed the structure of the optimal cooperation policy and showed that the policy is also optimal in terms of system throughput under some mild conditions. Based on the long-term payoff of each potential matching, we used the matching game with transfer to solve the pairing problem.  A distributed algorithm was proposed to find the $\epsilon$-stable matching. We investigated the optimality and complexity of the algorithm and proved that the algorithm is robust to the unilateral deviation of D2D pairs. Finally, the performance advantage of the proposed algorithm was confirmed by simulations. We end our conclusion by pointing out that there are some other techniques, e.g., coordinated multiple points (CoMP), which also could improve the performance of the cell-edge users and avoid frequent handover, and that the  combination of the proposed scheme and these techniques are actually  interesting and open research topics in the field.%will be investigated in the future.}

\section*{Appendix} 
\subsection{Proof of Proposition \ref{prop1}}
Consider the following problem
\begin{subequations}
\label{equ2-17}
\begin{alignat}{3}
  \max_{\pi}\quad &\mathbb{E}\left\{\pi(\mathbf{r}_{mn})r^D_{mn}+\eta \left(1-\pi(\mathbf{r}_{mn})\right)r^C_{mn}\right\}\\
  \text{s.t.}\quad & \mathbb{E}\left\{\left(1-\pi(\mathbf{r}_{mn})\right)r^C_{mn}\right\}\geq r_{th}.
  \end{alignat}
\end{subequations}
Thus, the Lagrangian of problem (\ref{equ2-17}) is represented as 
\begin{equation}
\label{equ2-18}
\begin{split}
\mathcal{L}'(\pi,\lambda')&=\mathbb{E}\left\{\pi(\mathbf{r}_{mn})(r^D_{mn}-(\eta+\lambda'))\right\}\\
&\qquad+(\eta+\lambda')\mathbb{E}\{r^C_{mn}\}-\lambda' r_{th}.
\end{split}
\end{equation}

\par Comparing (\ref{equ2-18}) with (\ref{equ2-9}), it can be found that if we let $\lambda=\lambda'+\eta$, the difference between $\mathcal{L}'(\pi,\lambda')$ and $\mathcal{L}(\pi,\lambda)$ is only a constant $\eta r_{th}$. Therefore, if $\lambda^*\geq\eta$, a similar argument as the proof of Theorem \ref{thm1} can be used to show that $\pi^*$ is the optimal solution to problem (\ref{equ2-17}). Note that $\mathbb{E}\left\{r^C_{mn}\mathbf{I}(\eta r^C_{mn}\geq r^D_{mn})\right\}<r_{th}$ is a sufficient condition of $\lambda^*\geq\eta$. In the following, we will show that this sufficient condition holds if (\ref{equ2-16}) is satisfied.

\par Based on the Cauchy inequality $\mathbb{E}\{XY\}\leq\sqrt{\mathbb{E}\{X^2\}\mathbb{E}\{Y^2}\}$, we have
\begin{equation}
\label{equ2-19}
\begin{split}
\mathbb{E}\left\{r^C_{mn}\mathbf{I}(\eta r^C_{mn}\geq r^D_{mn})\right\}
&\leq \sqrt{\mathbb{E}\left\{\mathbf{I}^2(\eta r^C_{mn}\geq r^D_{mn})\right\}}\\
&\qquad\times\sqrt{\mathbb{E}\left\{(r^C_{mn})^2\right\}}\\
&= \sqrt{\mathbb{E}\left\{\mathbf{I}(\eta r^C_{mn}\geq r^D_{mn})\right\}}\\
&\qquad\times\sqrt{\mathbb{E}\left\{(r^C_{mn})^2\right\}}\\
&= \sqrt{\text{Pr}\left\{\eta r^C_{mn}\geq r^D_{mn}\right\}\mathbb{E}\left\{(r^C_{mn})^2\right\}}.
\end{split}
\end{equation}

\par Note that (\ref{equ2-16}) can be converted as 
\begin{equation}
\label{equ2-20}
\text{Pr}\left\{\eta r^C_{mn}\geq r^D_{mn}\right\}\mathbb{E}\left\{(r^C_{mn})^2\right\}<r_{th}^2.
\end{equation}

\par Thus, we obtain $\mathbb{E}\left\{r^C_{mn}\mathbf{I}(\eta r^C_{mn}\geq r^D_{mn})\right\}<r_{th}$ if (\ref{equ2-20}) is plugged into (\ref{equ2-19}). Therefore, $\pi^*$ is the optimal solution to problem (\ref{equ2-17}).

\par Furthermore, if problem (\ref{equ2-15}) is feasible, then $\pi^*$ must be a feasible solution to problem (\ref{equ2-15}). Moreover, the optimal value of problem (\ref{equ2-17}) is the upper bound of the optimal value of problem (\ref{equ2-15}). Therefore, we conclude that $\pi^*$ is also the optimal solution to problem (\ref{equ2-15}).

\subsection{Proof of Lemma \ref{lemma2-2}}
To prove Lemma \ref{lemma2-2}, we will use the following lemma.
\begin{lemma}
\label{lemma2-1}
Denote the $\epsilon$-stable matching obtained by the DMA as $\Phi=(\mu,\mathbf{p})$. Then, there is no price vector $\mathbf{q}$ such that $(\mu,\mathbf{q})$ is $\epsilon$-stable and $\mathbf{q}$ satisfies
\begin{equation}
\label{equ2-29}
q_m=
\begin{cases}
p_m-2\epsilon,&\quad m\in\mathcal{C},\\
p_m,			&\quad m\in\mathcal{M}\setminus\mathcal{C},
\end{cases}
\end{equation}
where $\mathcal{C}\subseteq\mathcal{M}$.
\end{lemma}
\begin{IEEEproof}
We will prove the lemma by contradiction.

\par Suppose there exist a set $\mathcal{C}$ and a price vector $\mathbf{q}$ which satisfy (\ref{equ2-29}) and $\Phi'=(\mu,\mathbf{q})$ is $\epsilon$-stable. The set of the D2D pairs that are matched with the CU in $\mathcal{C}$ is denoted by $\mathcal{D}=\{n|\mu(n)\in\mathcal{C}\}$. Thus, for $\forall m\in\mathcal{C}$, when $\beta_m=q_m+\epsilon$ and $\beta_m=q_m+2\epsilon$ during the DMA, there must be a D2D pair proposing to or being matched with $CU_m$. The set of those D2D pairs is denoted by $\mathcal{D}_m$. Define $\mathcal{D}'\triangleq\cup_{m\in\mathcal{C}}\mathcal{D}_m$. Note that at least one D2D pair in $\mathcal{D}'$ proposes to or is matched with $CU_m$ once $\beta_m$ increases from $q_m+\epsilon$ to $q_m+2\epsilon$. Therefore, we must have $|\mathcal{D}'|>|\mathcal{D}|$ (otherwise, there exists a CU $m\in\mathcal{C}$ such that $\beta_m$ will not increase from $q_m+\epsilon$ to $q_m+2\epsilon$).

\par Since $\mathcal{D}'\setminus\mathcal{D}$ is nonempty, we assume there exist $n_1\in\mathcal{D}'\setminus\mathcal{D}$ and $m\in\mathcal{\mathcal{C}}$ such that $DP_{n_1}$ proposes to $CU_m$ when $\beta_m = q_m+\epsilon$ at some iteration of the DMA (the argument is similar for the cases where $\beta_m = q_m+2\epsilon$). Thus, we have
\begin{equation}
v_{mn_1}-(q_m+\epsilon)>v_{\mu(n_1)n_1-p_{\mu(n_1)}}.
\end{equation}
Therefore, for the matching $\Phi'=(\mu,\mathbf{q})$, we obtain
\begin{equation}
v_{mn_1}-\epsilon>v_{\mu(n_1)n_1}-p_{\mu(n_1)}+q_m=\delta_{n_1}(\Phi')+\theta_m(\Phi'), 
\end{equation}
where the equality is based on the fact that $p_{\mu(n_1)}=q_{\mu(n_1)}$ since $n_1\notin\mathcal{D}$.

\par Therefore, $\Phi'$ is not $\epsilon$-stable, which contradicts with our assumption.
\end{IEEEproof}

In the following, we will use Lemma  \ref{lemma2-1} to prove Lemma  \ref{lemma2-2}.

\par Denote the final matching obtained by the DMA is $\Phi=(\mu,\mathbf{p})$, and $\mathbf{X}=[x_{mn}]$ is the assignment associated with $\Phi$. It is easy to verify that the lemma holds for the unmatched D2D pairs in the matching $\mu$. Therefore, we focus on the matched D2D pairs.

\par Construct a directed graph $G(\Phi)$ for the matching $\Phi$ with the vertices $\mathcal{M}\cup\mathcal{N}$. There are two kinds of edges: if $CU_m$ and $DP_n$ are matched, then there is an edge from $DP_n$ to $CU_m$; if $v_{mn}-\epsilon\leq\delta_n(\Phi)+\theta_m(\Phi)\leq v_{mn}+\epsilon$ and  $CU_m$ and $DP_n$ are not matched, then there is an edge from $CU_m$ to $DP_n$.

\par Assume $CU_m$ has a positive price, i.e., $p_m>0$. Then, there must be a directed path that starts from $CU_m$ and ends at an unmatched D2D pair or at a CU with price less than {$2\epsilon$}. To prove this, suppose there is no such path. The sets  of D2D pairs and CUs, that can be reached from $CU_m$, are denoted by $\mathcal{S}$ and $\mathcal{T}$, respectively. Then $p_i\geq 2\epsilon$ for all $i\in\mathcal{T}$. Note that $\theta_i(\Phi)-2\epsilon+\delta_j(\Phi)\leq v_{ij}-\epsilon$ for $\forall i\in\mathcal{T}$ and $\forall j\in\mathcal{N}\setminus\mathcal{S}$. Thus, we can decrease $p_i$ by $2\epsilon$ for all $i\in\mathcal{T}$ and still have an $\epsilon$-stable matching, which contradicts with Lemma \ref{lemma2-1}. Furthermore, {if $p_i=\epsilon$ for all $i\in\mathcal{T}$ (since $p_i$ only can be the multiple of $\epsilon$ in DMA),} then it is easy to find out that we can decrease $p_i$ by $\epsilon$ for all $i\in\mathcal{T}$ and still have an $\epsilon$-stable matching. In this new matching, there exists a CU with price zero in $\mathcal{T}$. Besides, all the CUs in $\mathcal{T}$ is still reached by $CU_m$ in the directed graph associated with the new matching. Therefore, for the matching $\Phi=(\mu,p)$ and $CU_m$, there exists an $\epsilon$-stable matching $\Phi_m=(\mu,\mathbf{p}')$, which satisfies
\begin{equation}
\label{equ2-35}
p_i-\epsilon\leq p'_i \leq p_i,\forall i\in\mathcal{M},
\end{equation}
such that in the directed graph $G(\Phi_m)$, there must be a path starting from $CU_m$ and ending at an unmatched D2D pair or at a CU with price zero.

\par Consider a matched D2D pair $n_0$ and assume $DP_{n_0}$ is matched with $CU_{m_1}$. Then, we have a new $\epsilon$-stable matching $\Phi_{m_1}=(\mu,\mathbf{p}')$ satisfying (\ref{equ2-35}), and in the graph $G(\Phi_{m_1})$, there is a path starting from $CU_{m_1}$ and ending at an unmatched D2D pair or at a CU with price zero.

\par We focus on the first case (the argument is similar for the second case). Without loss of generality, we assume that the path is  $(CU_{m_1},DP_{n_1},\cdots,CU_{m_s},DP_{n_s},CU_k)$. Define $\mathcal{M}_s=\{m_1,\cdots,m_s\}$ and $\mathcal{N}_s=\{n_1,\cdots,n_s\}$. It is obvious that $|\mathcal{M}_s|=|\mathcal{N}_s|\leq C_1$. Construct a new matching $\Phi'=(\mu',\mathbf{p}')$, that assigns $CU_{m_1}$ to $DP_{n_1}$, $CU_{m_2}$ to $DP_{n_2}$,$\cdots$, $CU_{m_s}$ to $DP_{n_s}$, and that leaves $CU_k$ unmatched. Thus, for $\forall m\in\mathcal{M}$ and $\forall n\in\mathcal{N}$, we have
\begin{equation}
\begin{split}
\theta_m(\Phi')+\delta_n(\Phi')
&=\theta_m(\Phi_{m_1})+v_{\mu'(n)n}-p'_{\mu'(n)}\\
&\geq\theta_m(\Phi_{m_1})+\delta_n(\Phi_{m_1})-\epsilon\\
&\geq v_{mn}-2\epsilon.
\end{split}
\end{equation}
According to the construction of $\Phi'$, it turns out that $v_{\mu'(n)n}+\epsilon\geq\theta_{\mu'(n)}(\Phi_{m_1})+\delta_n(\Phi_{m_1})=p'_{\mu'(n)}+\delta_n(\Phi_{m_1})$. Thus, the first inequality can be derived. The second inequality is based on the fact that $\Phi_{m_1}$ is $\epsilon$-stable. Besides, for $\forall n\in\mathcal{N}_s$, it can be found that $\delta_n(\Phi')=v_{\mu'(n)n}-p'_{\mu'(n)}\geq\delta_n(\Phi_{m_1})-\epsilon\geq-\epsilon$. Therefore, if we define {$\tilde{\delta}_n(\Phi')\triangleq\delta_n(\Phi')+\epsilon\geq 0$}  for $\forall n \in \mathcal{N}_s$, then using a similar argument as the proof of Theorem \ref{the2-2}, we obtain
\begin{equation}
\label{equ2-37}
-3\epsilon C_1\leq\sum_{m\in\mathcal{M},\atop n\in\mathcal{N}\setminus\{n_0\}}x'_{mn}v_{mn}- V(\mathcal{M},\mathcal{N}\setminus\{n_0\})\leq 0,
\end{equation}
where $\mathbf{X}'=[x'_{mn}]$ is the associated assignment of the matching $\Phi'$.

\par On the one hand, for $\forall n \in \mathcal{N}_s$, the following inequality holds
\begin{equation}
\label{equ2-38}
\delta_n(\Phi')=v_{\mu'(n)n}-\theta_{\mu'(n)}(\Phi_{m_1})\leq\delta_n(\Phi_{m_1})+\epsilon,
\end{equation} 
where the last inequality follows the fact that $\Phi_{m_1}$ is $\epsilon$-stable. Note that $|\mathcal{M}_s|=|\mathcal{N}_s|\leq C_1$. Therefore, combining this fact with (\ref{equ2-38}), we have
\begin{equation}
\label{equ2-39}
\begin{split}
\sum_{m\in\mathcal{M},\atop n\in\mathcal{N}\setminus\{n_0\}}\!\!\!\!\!\!x'_{mn}v_{mn}
&={\sum_{n \in {{\cal N}_s}} {{\delta _n}(\Phi ')} } + \!\!\!\!\!\!\!\!\!\!\!\!\sum_{n \in {\cal N}\setminus (\{ {n_0}\}  \cup {{\cal N}_s})} {{\!\!\!\!\!\!\!\!\!\!\!\!\delta _n}(\Phi ')}  + \sum_{m \in {\cal M}} {{\theta _m}(\Phi ')}\\
&\leq \sum_{n\in\mathcal{N}\setminus\{n_0\}}\!\!\!\!\!\!\delta_{n}(\Phi_{m_1}) +\sum_{m\in\mathcal{M}}\theta_m(\Phi_{m_1})+C_1\epsilon\\
&= \sum_{m\in\mathcal{M}, n\in\mathcal{N}}\!\!\!\!\!\!x_{mn}v_{mn} -\delta_{n_0}(\Phi_{m_1})+C_1\epsilon\\
&\overset{(a)}{\leq} \sum_{m\in\mathcal{M}, n\in\mathcal{N}}\!\!\!\!\!\!x_{mn}v_{mn} -\delta_{n_0}(\Phi)+C_1\epsilon\\
&\leq V(\mathcal{M},\mathcal{N})-\delta_{n_0}(\Phi)+C_1\epsilon,
\end{split}
\end{equation}
where inequality (a) is based on the fact that $\delta_{n_0}(\Phi)\leq\delta_{n_0}(\Phi_{m_1})$.

\par Furthermore, using (\ref{equ2-37}) and (\ref{equ2-39}), we get
\begin{equation}
V(\mathcal{M},\mathcal{N}\setminus\{n_0\})-3C_1\epsilon\leq V(\mathcal{M},\mathcal{N})-\delta_{n_0}(\Phi)+C_1\epsilon.
\end{equation}
Therefore, we conclude that  (\ref{equ2-33}) holds.

\par On the other hand, using $\delta_n(\Phi')=v_{\mu'(n)n}-p'_{\mu'(n)}\geq\delta_n(\Phi_{m_1})-\epsilon\geq$ and $|\mathcal{M}_s|=|\mathcal{N}_s|\leq C_1$, we obtain
\begin{equation}
\label{equ2-40}
\begin{split}
\sum_{m\in\mathcal{M},\atop n\in\mathcal{N}\setminus\{n_0\}}\!\!\!\!\!\!x'_{mn}v_{mn}
&={\sum_{n \in {{\cal N}_s}} {{\delta _n}(\Phi ')} } + \!\!\!\!\!\!\!\!\!\!\!\!\sum_{n \in {\cal N}\setminus (\{ {n_0}\}  \cup {{\cal N}_s})} {{\!\!\!\!\!\!\!\!\!\!\!\!\delta _n}(\Phi ')}  + \sum_{m \in {\cal M}} {{\theta _m}(\Phi ')}\\
&\geq \sum_{n\in\mathcal{N}\setminus\{n_0\}}\!\!\!\!\!\!\delta_{n}(\Phi_{m_1}) +\sum_{m\in\mathcal{M}}\theta_m(\Phi_{m_1})-C_1\epsilon\\
&= \sum_{m\in\mathcal{M}, n\in\mathcal{N}}\!\!\!\!\!\!x_{mn}v_{mn} -\delta_{n_0}(\Phi_{m_1})-C_1\epsilon\\
&\overset{(a)}{\geq} \sum_{m\in\mathcal{M}, n\in\mathcal{N}}\!\!\!\!\!\!x_{mn}v_{mn} -\delta_{n_0}(\Phi)-C_1\epsilon-\epsilon\\
&\overset{(b)}{\geq} V(\mathcal{M},\mathcal{N})-\delta_{n_0}(\Phi)-(C_1+C_2+1)\epsilon,
\end{split}
\end{equation}
where inequality (a) is from the fact that $\delta_{n_0}(\Phi)+\epsilon\geq\delta_{n_0}(\Phi_{m_1})$, and inequality (b) is based on Theorem \ref{the2-2}. Furthermore, combining (\ref{equ2-40}) with (\ref{equ2-37}), we conclude that (\ref{equ2-32}) holds.

\bibliographystyle{IEEEtran}
\bibliography{IEEEabrv, ref.bib}
\end{document}